\documentclass[11pt]{article}
\usepackage{amsmath,amssymb,amsfonts,graphicx}
\usepackage{jheppub}
\usepackage{subcaption}
\usepackage{comment}
\usepackage{xcolor}

\newcommand{\be}{\begin{equation}}
\newcommand{\ee}{\end{equation}}
\newcommand{\bea}{\begin{eqnarray}}
\newcommand{\eea}{\end{eqnarray}}
\newcommand{\beas}{\begin{eqnarray*}}
\newcommand{\eeas}{\end{eqnarray*}}
\newcommand{\ba}{\begin{array}}
\newcommand{\ea}{\end{array}}
\newcommand{\tr}{{\rm tr}}

\newcommand{\rd}{\mathrm{d}}

\newcommand{\ket}[1]{\left| #1 \right\rangle}

\newcommand{\avg}[1]{\left\langle #1 \right\rangle}

\title{Menagerie of Euclidean constructions for 3D holographic cosmologies}

\author{Mark Van Raamsdonk,}
\author{Alejandro Vilar L\'opez}
\affiliation{Department of Physics and Astronomy, University of British Columbia,
6224 Agricultural Road, Vancouver, B.C.\ V6T 1Z1, Canada}

\emailAdd{mav@phas.ubc.ca}
\emailAdd{alejandro.vilarlopez@ubc.ca}

\abstract{We construct a large number of exact solutions of three-dimensional gravity with heavy matter particles that generalize the construction of Antonini, Sasieta, and Swingle (AS${}^2$), argued to define CFT states dual to a spacetime with a closed baby universe cosmology. Our construction starts with an arbitrary heavy-particle closed universe cosmology of the type constructed in \cite{Maloney:2025tnn}, and via a gluing procedure adds an arbitrary number of AdS tubes connecting the past and future conformal boundaries of the associated Euclidean wormhole solution. With our construction, it is straightforward to produce examples where the cosmology is approximately homogeneous and isotropic. We describe a necessary condition for the cosmological wormhole saddle to dominate the Euclidean path integral with the specified boundary conditions. We argue that the original AS${}^2$ construction usually does not meet this condition, and describe alternative saddles that are likely to dominate. We discuss various possibilities for how the cosmological saddle might be made to dominate in our generalized construction.}

\begin{document}

\maketitle

\section{Introduction}

Recently, there has been significant interest in the description of closed universe cosmologies in quantum gravity. Some discussions make use of formal arguments involving the gravitational path integral; these have suggested that the Hilbert space for a closed universe should be one-dimensional \cite{Penington:2019kki,Marolf:2020xie,Usatyuk:2024mzs,Usatyuk:2024isz}\footnote{Other arguments that hint towards essentially the same conclusion can be found in \cite{Maldacena:2004rf,Almheiri:2019hni,McNamara:2020uza}.} and led some authors to speculate that observers must be introduced explicitly in the formalism in order to recover a more conventional picture of the physics with a non-trivial Hilbert space \cite{Abdalla:2025gzn,Harlow:2025pvj,Akers:2025ahe,Chen:2025fwp}. A different approach has been to construct closed universe cosmologies in the context of negative $\Lambda$ gravitational effective theories and to try to interpret these solutions microscopically using the tools of holography \cite{Maldacena:2004rf,McInnes:2004nx,Banerjee:2018qey,Antonini:2019qkt,VanRaamsdonk:2021qgv,Sahu:2024ccg,Maloney:2025tnn}. In this approach, it has been argued generally in \cite{VanRaamsdonk:2020tlr} that in the holographic description of a closed universe cosmology we generally find a set of auxiliary degrees of freedom that provide a Hilbert space for the cosmology. This is reminiscent of the observer story, but here the auxiliary degrees of freedom are built in to the construction, as opposed to being added. 

In \cite{Antonini:2023hdh} (which we will refer to as AS${}^2$), Antonini, Sasieta, and Swingle present an intriguing Euclidean path integral construction for a state of two CFTs on $S^d$ that is proposed to be dual to a pair of AdS spacetimes together with a disconnected closed universe cosmology. The cosmology is an anisotropic big bang / big crunch negative $\Lambda$ cosmology supported by a shell of matter distributed on the equator of a spherical topology universe that is otherwise empty. There has been significant debate in the literature about whether the CFT state constructed by \cite{Antonini:2023hdh} actually encodes the physics of the cosmology. In \cite{Antonini:2024mci}, the authors pointed out that the standard AdS/CFT dictionary would interpret the ${\rm AS}^2$ CFT state as dual to two AdS regions containing an entangled state of low-energy gases of particles. This would lead to the puzzling conclusion that a single CFT state has two dual gravitational interpretations: one with a closed universe and one without it. Arguments against the semiclassical validity of the closed universe picture were given in \cite{Engelhardt:2025vsp,Gesteau:2025obm}, while \cite{Antonini:2025ioh} argued that the breakdown of semiclassicality is not necessarily severe, or even inexistent if a notion of averaging is introduced. From a different perspective, \cite{Kudler-Flam:2025cki} showed that, in situations with low entanglement (not scaling with $N$), the dominance of the saddle containing a disconnected closed universe implies that the state does not have a good large-$N$ limit, calling into question the status of the construction.\footnote{In the context of two-dimensional gravity, a model that avoids some of these issues was introduced in \cite{Sasieta:2025vck}.} Alternatives to interpret the cosmological saddle in terms of a modified, averaged large-$N$ limit have been proposed \cite{Kudler-Flam:2025cki,Liu:2025cml,Liu:2025ikq}, but there is currently no fully satisfactory implementation of this idea.

In this paper, our goal is to extend and comment on the AS${}^2$ construction in several ways, adding to the previous discussion another possibility for interpreting the ${\rm AS}^2$ saddle -- namely, that it is subdominant.\footnote{The recent work \cite{Belin:2025ako} also reaches this conclusion for a special case of our construction.} First, using the methods of \cite{Maloney:2025tnn}, we construct a large class of fully back-reacted three-dimensional gravity solutions that realize the AS${}^2$ construction at a higher level of microscopic detail and significantly generalize it. While the AS${}^2$ construction approximates the matter as a uniform shell (or ring in the 3D case), our construction treats each matter particle individually as a gravitational source producing a conical defect. This applies for matter particles with masses of order $c / \ell_{AdS} \sim  1/ G$ but less than the black hole threshold $m_* = 1/(4G)$. These particles correspond to individual operator insertions in the corresponding CFT path integral. Our generalization allows cosmologies with general spatial topology and completely general distributions of matter particles (satisfying appropriate constraints). In particular, it allows the construction of cosmologies that are arbitrarily close to being homogeneous and isotropic. The techniques we develop allow also to treat generalizations of the ${\rm AS}^2$ setup in which the cosmology is empty and the spatial slices are surfaces with genus $g \geq 2$, including as a special case the model discussed in \cite{Belin:2025ako}.

\begin{figure}
    \centering
\includegraphics[width=0.7\linewidth]{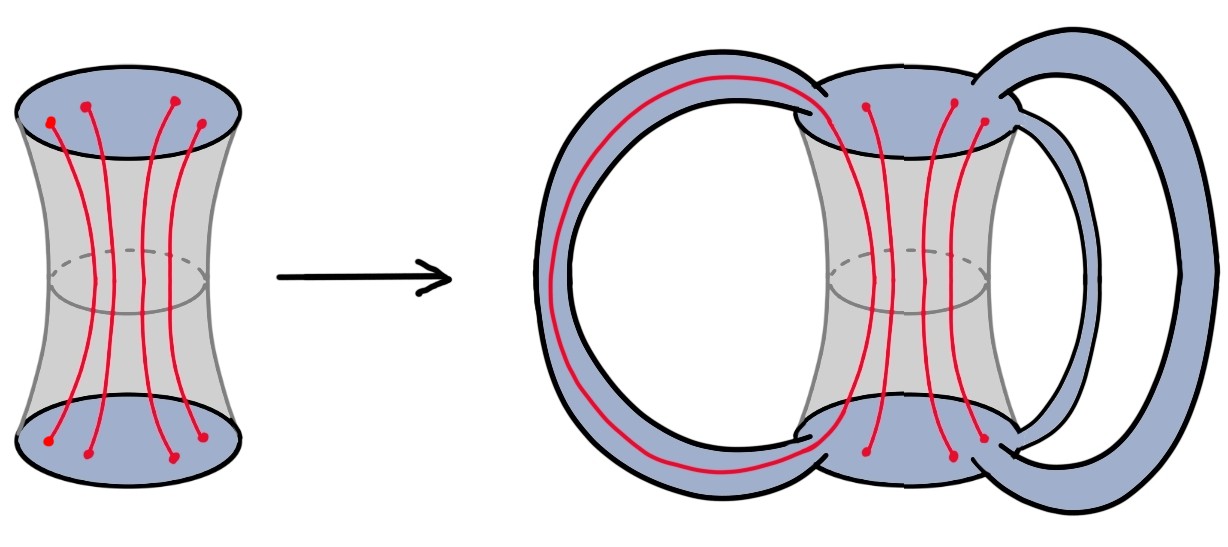}
    \caption{Left: A Euclidean wormhole supported by matter particles corresponding to operator insertions; the Lorentzian  continuation is a closed universe big-bang / big-crunch cosmology. Right: via surgery, we can add  arbitrary numbers of AdS cylinders connecting the past and future conformal boundaries. These may include some of the matter particles (removing the portion of the boundary that contained the corresponding operator  insertion). The Lorentzian continuation now contains the cosmology and a number of asymptotically AdS spacetimes, some containing partner particles entangled with particles in the cosmology.}
    \label{fig:IntroPic}
\end{figure}

We show that for a given cosmology, there is actually an infinite family of possible Euclidean constructions that continue to a Lorentzian spacetime including the cosmology. Each of these is related to a parent Euclidean wormhole of the type constructed in \cite{Maloney:2025tnn}, which provides the simplest analytic continuation of the cosmology. The associated AS${}^2$ constructions are obtained from this parent wormhole via a surgery that connects the two sides of the wormhole by gluing in additional finite AdS cylinders, or more generally, AdS cylinders with matter particles. This is illustrated in Figure \ref{fig:IntroPic}. The various gluings preserve the time-reflection symmetry of the wormhole. For each additional cylinder that we glue in, the associated Lorentzian spacetime contains an additional copy of AdS entangled with the cosmology. 

With our construction, there is now a huge variety of possible CFT states whose dual potentially includes the cosmology. A key question is whether any of these have the cosmology as the dominant part of the dual gravitational wavefunction. This requires that the associated Euclidean geometry provides the dominant saddle with the specified boundary conditions. In section \ref{sec:condition}, we introduce a necessary condition for this dominance (previously mentioned briefly in \cite{Sahu:2024ccg}). The idea is that if we have a CFT/gravity path integral whose dominant saddle contains a wormhole that continues to a cosmology, slicing this path integral along the length of the wormhole constructs a state that contains a Lorentzian wormhole. The dual state constructed by the CFT version of this sliced path integral must be an  entangled state of two CFTs, and by the Ryu-Takayanagi formula, the entanglement entropy must be of order $c$. Requiring the path integral construction to produce entanglement entropy of this order is our necessary condition.

Applying our criterion to the original AS${}^2$ construction, we find that the necessary condition does not appear to be satisfied, or at least would require extreme fine tuning in the choice of operator insertions in order to be satisfied. We discuss in section \ref{sec:othersaddles} other saddles that could dominate in our version of the AS${}^2$ construction, providing explicit calculations showing that these have lower action.

In section \ref{sec:othersaddles}, we also ask whether any of our more general path integral constructions could give a dominant saddle containing the cosmology. There is already one such example, discussed by one of us in \cite{Sahu:2024ccg}, where the ``particles'' in the cosmology are taken to have masses sufficiently above the black hole threshold so that they become relatively large black holes. More generally, we argue that a dominant saddle might arise in the situation where (at some large but finite $c$), we have of order $c$ additional tubes glued in to connect the two sides of the cosmological wormhole. In this case, the dual state would be an entangled state of a very large number of CFTs. We suggest one other possibility for cosmological dominance in the discussion: in certain cases, it may be that while the action of any individual cosmological saddle is larger that that of certain non-cosmological saddles, there are a vast number of possible cosmological saddles that result in saddles of cosmological type dominating the Euclidean path integral and thus the Lorentzian wavefunction.

\section{Setup and Generalities}
\label{sec:setup}

In this paper, we describe a large class of pairs $(M,M_E)$ of solutions of three-dimensional gravity with $\Lambda < 0$, where $M_E$ is a connected Euclidean spacetime with time-reflection symmetry and $M$ is a Lorentzian continuation of $M_E$ which includes one or more closed universe cosmologies. The solutions are all locally AdS${}_3$ with conical defects and will be constructed using the tools in \cite{Sahu:2024ccg,Maloney:2025tnn} that we now briefly review. 

\subsection{Cosmological wormholes}

The starting point for our construction is a special case described in \cite{Maloney:2025tnn} where $M_E$ is a time-reflection symmetric Euclidean wormhole solution and $M$ is a closed universe big-bang / big-crunch cosmology.

\begin{figure}
 \centering
    \includegraphics[scale = 0.3]{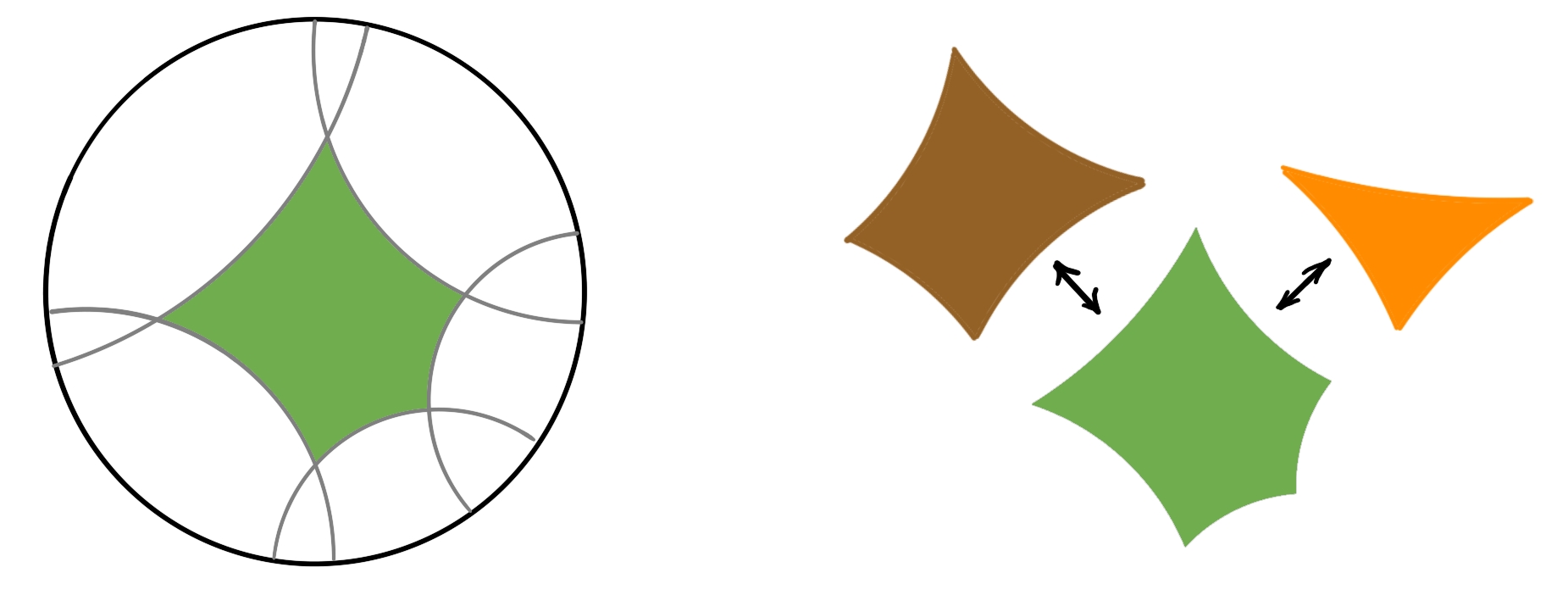}
    \caption{Left: a geodesic polygon in $H^2$. Right: Polygons with matching side lengths can be glued together smoothly, with possible conical defects at the vertices.}
    \label{fig:gluing}
\end{figure}

The spatial geometry $\Sigma$ of the slices fixed by time-reflection symmetry is common to $M$ and $M_E$. It is a compact space obtained by gluing together geodesic polygons, as illustrated in Figure \ref{fig:gluing}.

When gluing together polygons, we can smoothly glue any two sides with equal geodesic length; the sides may be from two different polygons or two different sides of the same polygon, however we assume that all the polygons are oriented and that the gluings respect orientation. We consider gluing a collection of polygons such that the final surface is compact and without boundaries. The resulting space may include conical defects at the vertices. These correspond to the locations of massive particles. For a conical deficit $\delta$, we require a particle of mass
\begin{equation}
\label{DeltaM}
8 \pi Gm = \delta
\end{equation}
in order that Einstein's equations are satisfied. We restrict to the situation with no conical excesses so that all particles have positive mass.\footnote{See \cite{Maloney:2025tnn} for a description of the most general such compact space.}

Given the geometry $\Sigma$ with metric $\rd \Sigma^2$, the Euclidean wormhole geometry $M_E$ can be described most simply via the metric
\begin{equation}
\rd s^2 = \rd u^2 + \cosh^2(u/\ell) \, \rd \Sigma^2 \, .
\label{Wormhole}
\end{equation}
This has two conformal boundaries, each with the same conformal geometry as $\Sigma$, at $u = \pm \infty$.

\begin{figure}
 \centering
    \includegraphics[scale = 0.5]{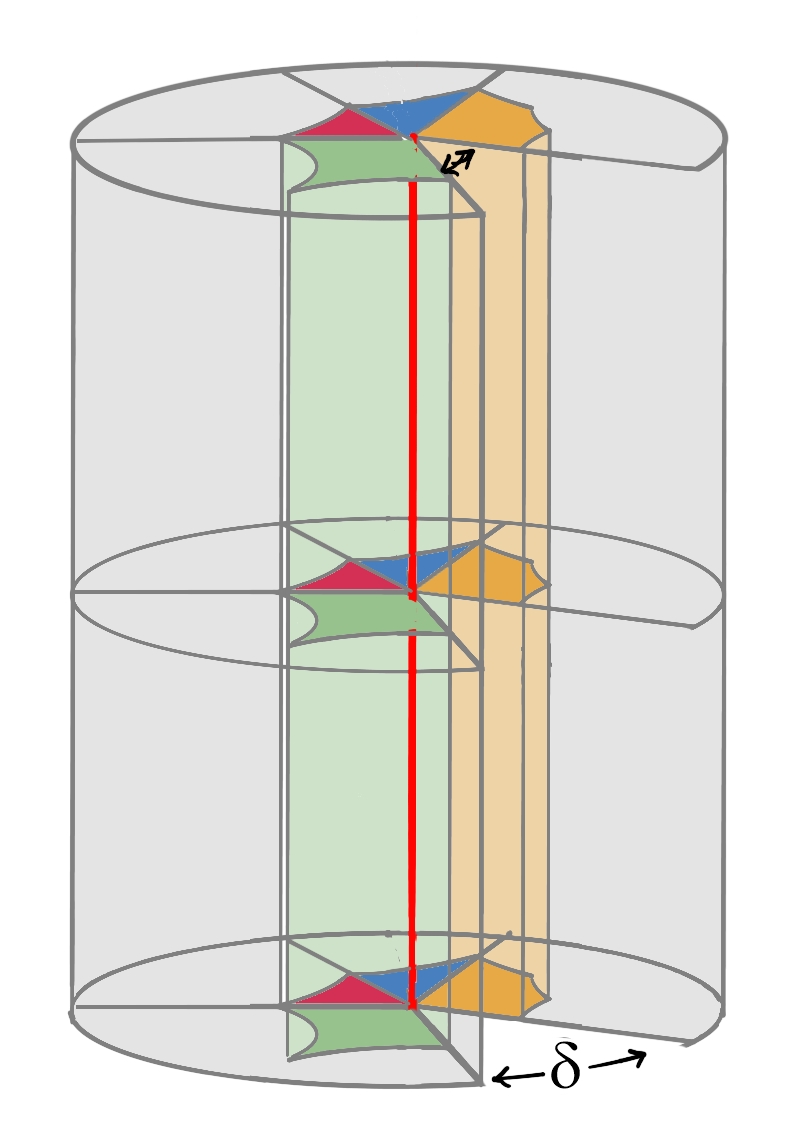}
    \caption{Regions adjacent to a conical defect (red) can be represented as geodesic prisms in FRW-coordinates. Note that this picture is not the usual AdS cylinder with time-translation symmetry; the conformal boundary is at $u = \pm \infty$ in these coordinates. }
    \label{fig:FRW}
\end{figure}

The associated cosmology $M$ is described by the analytic continuation
\begin{equation}
\rd s^2 = -\rd  t^2 + \cos^2(t/\ell) \, \rd \Sigma^2
\end{equation}
These geometries are locally AdS${}_3$ away from the conical singularities. The geometry near any one of the conical defects can be represented as in Figure \ref{fig:FRW}; the time-evolution of each polygon adjacent to the vertex can be represented as a geodesic prism in FRW coordinates with the chosen vertex at the center of the hyperbolic disk.

There is a nice relation between the total amount of matter in the cosmology, the 2D spatial volume and the genus given by  \cite{Maloney:2025tnn}
\begin{equation}
\label{Mass}
    M_{\rm tot} = \frac{1}{4 G} \left( \frac{V}{2 \pi \ell^2} + \chi \right) \; .
\end{equation}

\subsubsection*{Alternative representation}

For our construction below, it will be useful to give an alternative representation of the wormhole spacetimes.

For each polygon in $\Sigma$, consider a copy of Euclidean AdS described by the metric 
\begin{equation}
\label{cylinder}
\rd s^2 = \ell^2\left[
\left(\frac{1+x^2+y^2}{\,1-x^2-y^2\,}\right)^{\!2}\rd \tau^2
+\frac{4}{(1-x^2-y^2)^2}\,(\rd x^2+\rd y^2)
\right] \, . 
\end{equation}
Choose an embedding of the polygon in the $\tau = 0$ slice such that $x=y=0$ is an interior point of the polygon. In these coordinates, each side of the polygon is a geodesic described by the arc of a circle $C_i$ that intersects the boundary circle $x^2 + y^2 =1$ orthogonally. Each such circle is the $\tau=0$ slice of a unique time-reflection invariant two-dimensional geodesic surface $S_i$ in AdS${}_3$.\footnote{In Poincar\'e coordinates 
\begin{equation}
\rd s^2 = \frac{\ell^2}{Z^2}\left( \rd Z^2 + \rd X^2 + \rd Y^2 \right) \, ,
\end{equation}
where $\Sigma$ is the surface $Y=0$ and $C_i$ is the arc with $Z^2 + X^2 = 1$, the surface $S_i$ is described by $Z^2 + X^2 + Y^2 = 1$. In the cylinder coordinates AdS${}_3$, $S_i$ intersects the boundary of the cylinder on a closed curve symmetric under horizontal and vertical reflections.}

\begin{figure}
 \centering
    \includegraphics[width = 0.7 \linewidth]{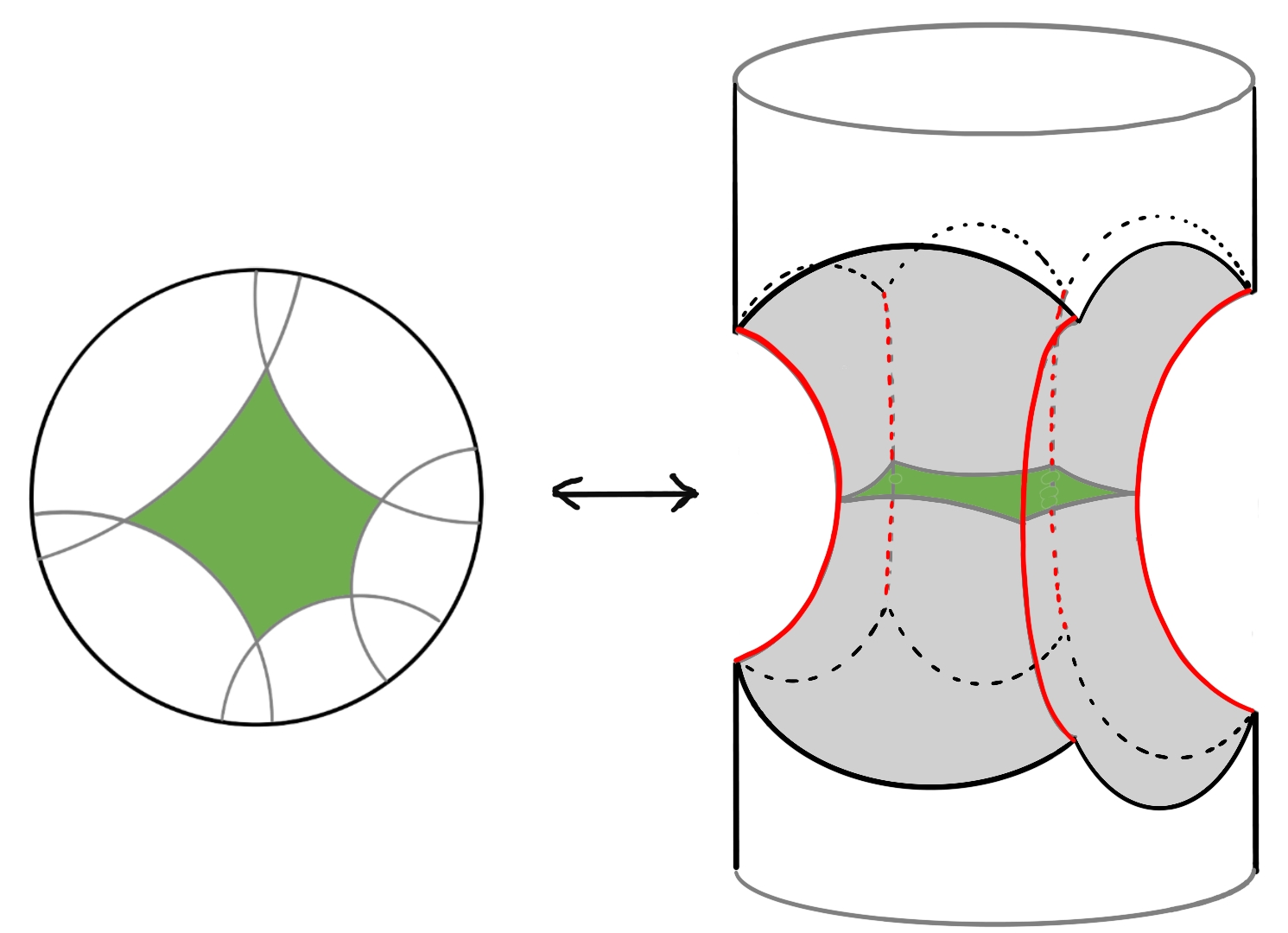}
    \caption{Each hyperbolic polygon $P$ in the Poincar\'e disk can be extended to a time reflection symmetric region $V_P$ of ${\rm AdS}_3$ bounded by geodesic surfaces. In the coordinates \eqref{cylinder}, if the center of the hyperbolic disk $\tau=0$ is an interior point of the polygon, $V_P$ contains the two asymptotic regions $\tau = \pm \infty$. We refer to such a $V_P$ as a chewed AdS${}_3$.}
    \label{fig:chewed}
\end{figure}

We extend our polygon $P$ to a region $V_P$ of AdS${}_3$ obtained by removing the exterior of each surface $S_i$ (i.e., the side not including the polygon). In the cylinder coordinates, the remaining portion of AdS${}_3$ is reminiscent of a tree trunk that has been chewed by a beaver. We will refer to $V_P$  as a ``chewed AdS${}_3$''. The geometry $V_P$ corresponding to polygon $P$ is illustrated in Figure \ref{fig:chewed}.

\begin{figure}
 \centering
    \includegraphics[width= 0.8 \linewidth]{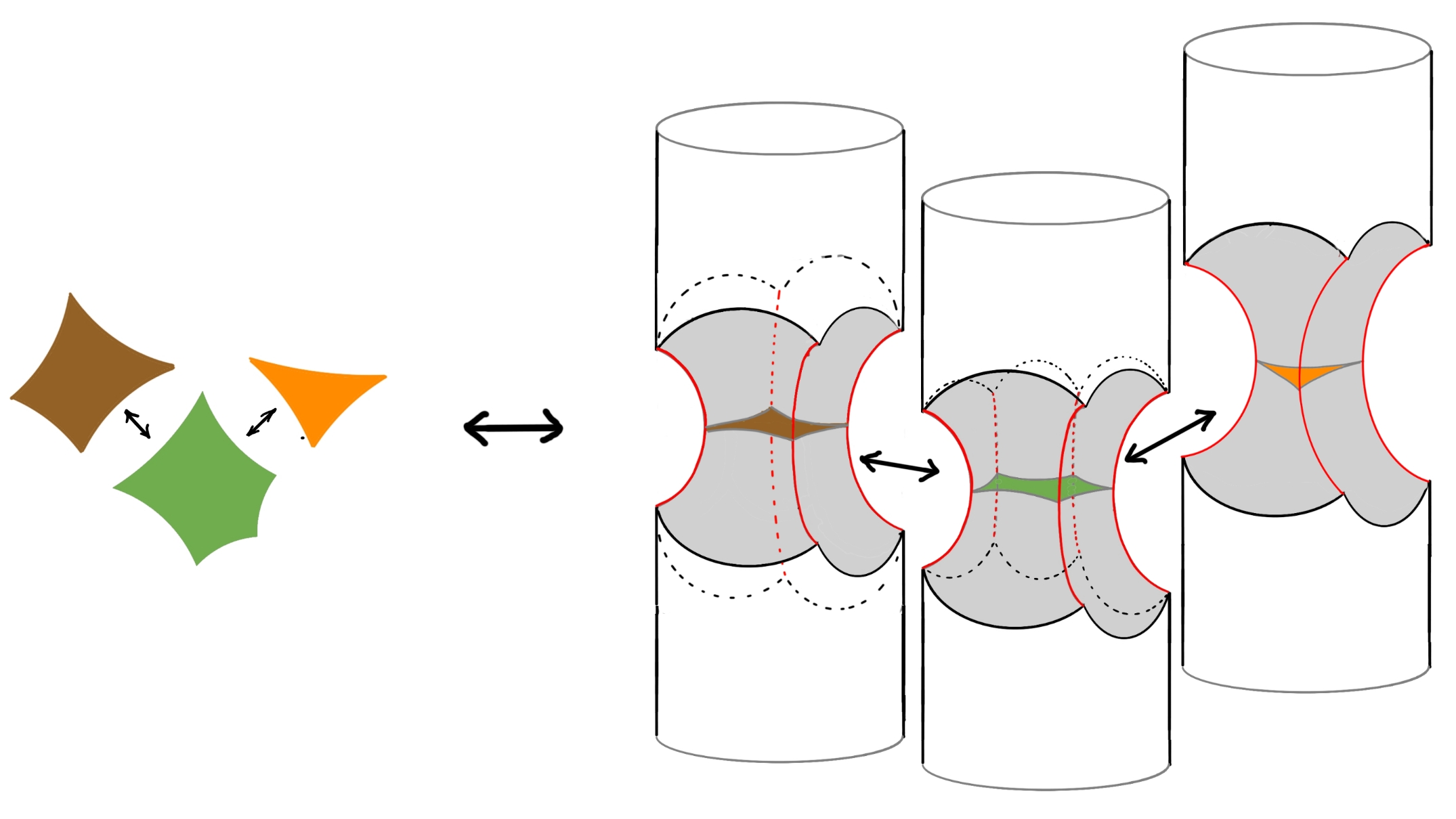}
    \caption{Each gluing of hyperbolic polygons to form a closed surface can be associated with a gluing of chewed AdS spaces to describe a wormhole geometry.}
    \label{fig:GluedChewed}
\end{figure}

For a closed surface $\Sigma$ obtained by gluing together polygons $\{P_n\}$, we can describe a time reflection symmetric locally AdS${}_3$ geometry for which this surface is the the time-reflection invariant slice by simply gluing together the set of chewed AdS geometries $\{V_{P_n}\}$ along the surfaces $S_i$ corresponding to the glued polygon sides, as shown in Figure \ref{fig:GluedChewed}. This gives an alternative construction of the wormhole geometry that we described above via the metric \eqref{Wormhole}. Each prism in the FRW coordinates of Figure \ref{fig:FRW} corresponds to one $V_P$ in the static coordinates of Figure \ref{fig:chewed}.

\subsection{AS${}^2$ geometries}

Starting from one of the wormhole solutions that we have constructed in the previous section, we can construct a very large family of additional solutions whose Lorentzian continuation includes the same cosmology by a gluing procedure that we now describe. These new solutions generalize the AS${}^2$ construction in \cite{Antonini:2023hdh}.

We give two equivalent descriptions of the construction. First, starting with the representation of Figure \ref{fig:chewed}, we can obtain new solutions by truncating the top and bottom of one or more of the cylinders in a reflection-symmetric way and gluing them together. For each AdS, we have a continuous parameter that controls the amount of the cylinder that we keep. We could also choose to glue together the top and bottom cylinders after some permutation, though only certain choices here will respect the time-reflection symmetry. Together with the parameters associated with embedding each polygon in hyperbolic space, we have all together a large number of parameters characterizing the Euclidean solution when the number of polygons is large.

For each of these glued solutions, the slice left invariant by time-reflection symmetry now includes $n$ copies of $H^2$ (the gluing surfaces) when we have truncated and glued $n$ of the cylinders.  In the Lorentzian continuation, we have $n$ copies of Lorentzian AdS together with a single closed universe cosmology.\footnote{See the discussion for a generalization where we have arbitrary numbers of closed universes.} We will see below that the original AS${}^2$ construction represents a special case where $n=2$ and we start with a very particular wormhole.

\begin{figure}
    \centering
    \includegraphics[width = 0.9 \linewidth]{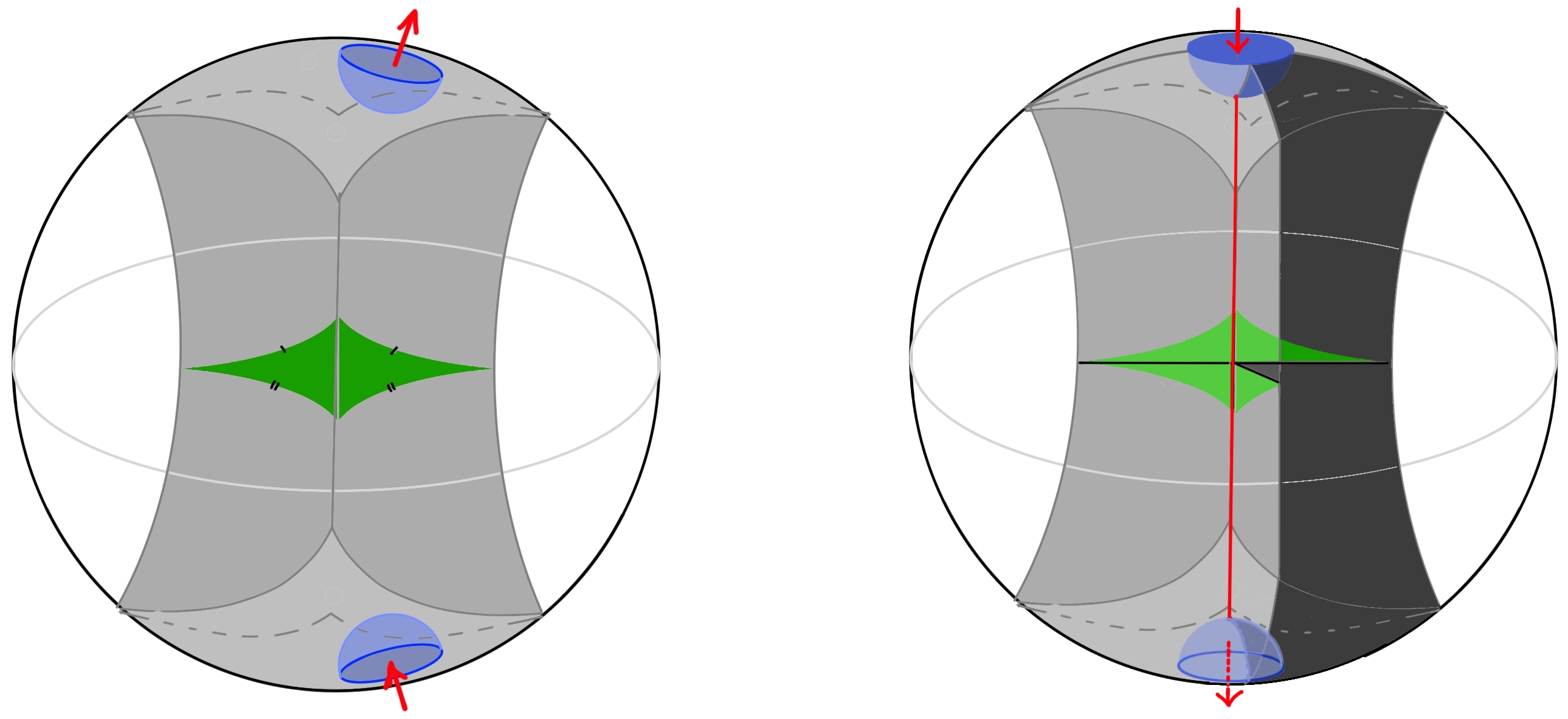}
    \caption{Left: Constructing an AS${}^2$-type geometry from a wormhole geometry by removing a region bounded by geodesic surface and its $\mathbb{Z}_2$ image and gluing across these surfaces. We can add arbitrarily many of these connections. Right: Gluing across a surface that includes a particle.}
    \label{fig:glueboth}
\end{figure}

An alternative way to understand the AS${}^2$-type solutions is to note that the wormhole geometry for a given $\Sigma$ has a pair of conformal boundaries related by time-reflection symmetry. Consider any extremal surface that ends on one of these conformal boundaries and avoids any of the conical singularities (particles). Consider also the image of this surface under the time-reflection symmetry. We can excise the part of the geometry between each of these surfaces and the conformal boundaries and then identify the two surfaces (or glue in a tube of AdS${}_3$ that connects them), as shown in Figure \ref{fig:glueboth} (left). In this way, we can add an arbitrary number of additional AdS${}_3$ ``tubes'' to the geometry while preserving the $\mathbb{Z}_2$ symmetry. 

We can perform a similar construction but include one or more of the particles in the glued tube region, as shown in Figure \ref{fig:glueboth} (right). In this case, we have an additional asymptotically AdS geometry in the Lorentzian solution that includes the particles that intersect the gluing surface.

The gluing picture suggests a generalization to arbitrary dimensions. Starting from any Euclidean wormhole solution that gives rise to a cosmology under analytic continuation, the asymptotic regions $\tau \to \pm \infty$ become arbitrarily close to pure AdS (since all contributions in the Euclidean Friedmann equation dilute faster than the negative cosmological constant as the scale factor grows). Thus, we can perform a similar gluing procedure, removing regions bounded by geodesic surfaces at either end of the wormhole and then identifying the solution across the two surfaces. 

\subsection{The AS${}^2$ construction with discrete particles}

\begin{figure}
    \centering
    \includegraphics[width = 0.5 \linewidth]{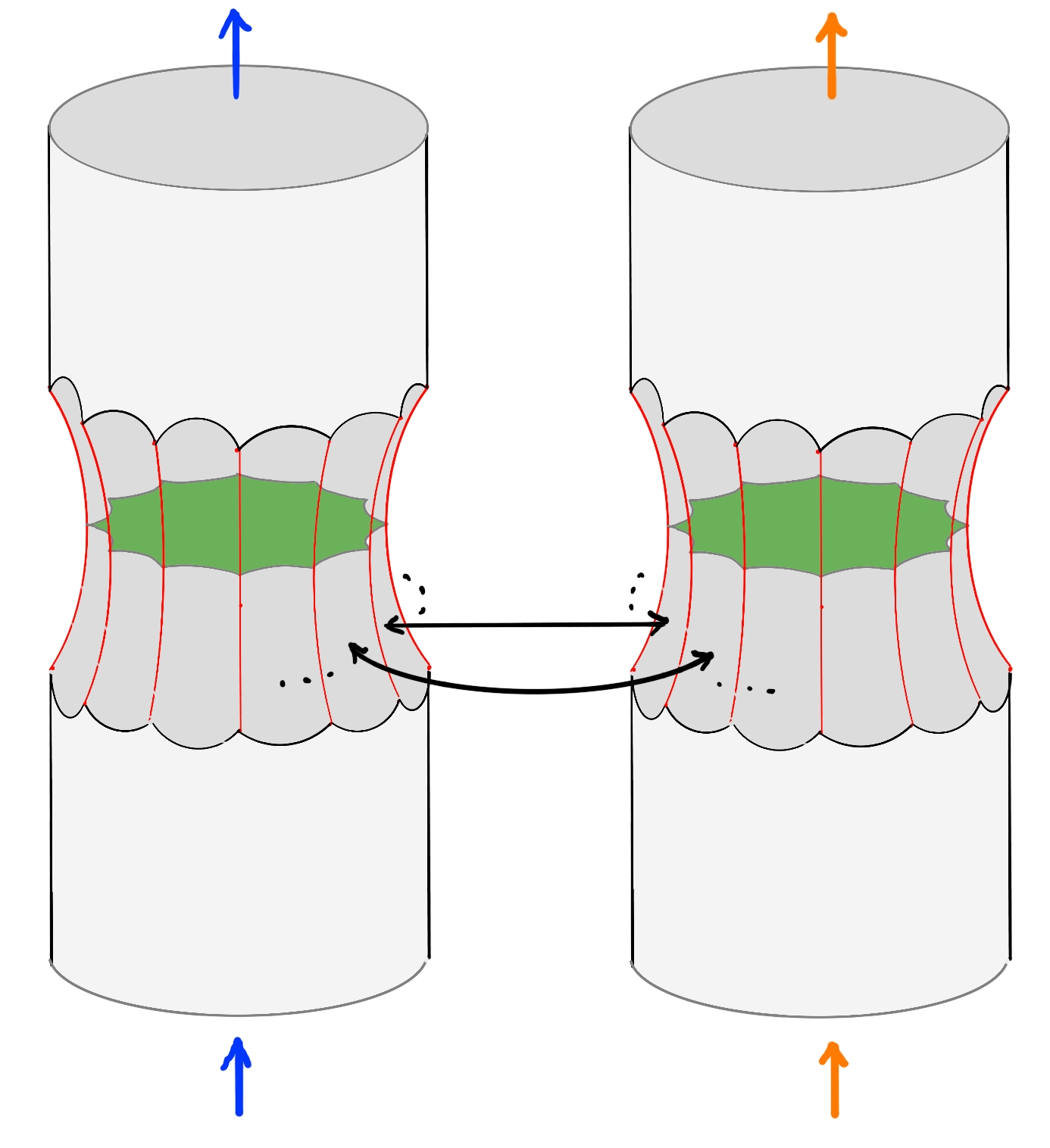}
    \caption{An exact discrete particle version of the original AS${}^2$ construction in AdS${}_3$.}
    \label{fig:OG}
\end{figure}

We can describe a class of solutions that matches precisely with the original AS${}^2$ construction as follows. Take any two regular hyperbolic $n$-gons for any $n \ge 3$ and glue sides cyclically, moving clockwise around one $n$-gon and counterclockwise around the other $n$-gon. The AS${}^2$-construction based on this gluing is depicted in Figure \ref{fig:OG}; we will see that it corresponds to a fully back-reacted unsmeared version of the original construction. 

\begin{figure}
    \centering
    \includegraphics[scale=0.35]{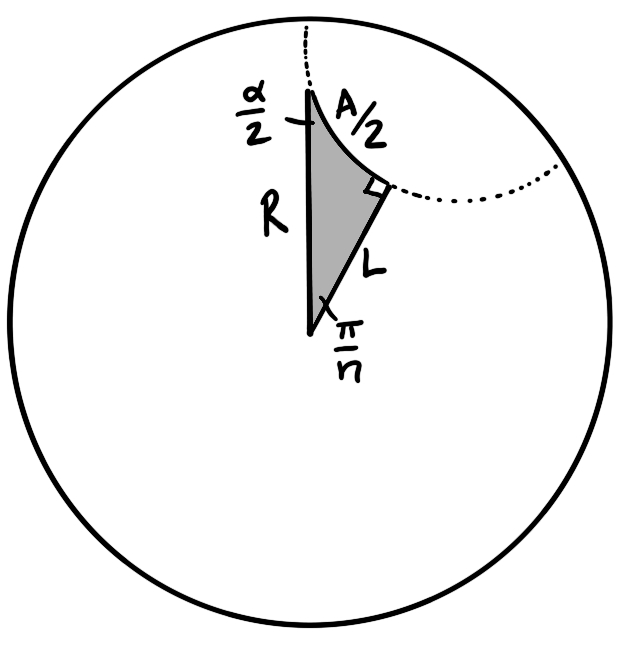}
    \caption{A hyerbolic $n$-gon with side length $A$ and vertex angle $\alpha$ can be decomposed into $2n$ of these right triangles.}
    \label{fig:triangle}
\end{figure}

The polygon can be decomposed into $2n$ right triangles of the type shown in Figure \ref{fig:triangle}.  Using the hyperbolic law of cosines, and momentarily setting $\ell=1$, we have that
\begin{equation}
\cosh R = \cosh L \cosh \frac{A}{2} \; .
\end{equation}
The hyperbolic sine law gives
\begin{equation}
\sinh R = \frac{\sinh L}{\sin (\alpha/2)} = \frac{\sinh (A/2)}{\sin (\pi/n)} \, .
\end{equation}
Squaring the cosine relation and using the sine law to eliminate $\sinh L$ and $\sinh(A/2)$, we get
\begin{equation}
1 = \sinh^2 R \, \sin^2 \frac{\alpha}{2} \, \sin^2 \frac{\pi}{n} + \sin^2 \frac{\alpha}{2} + \sin^2 \frac{\pi}{n} \; .
\end{equation}
The conical deficit is $\delta = 2 \pi - 2 \alpha$, so using \eqref{Mass}, the mass of each particle is
\begin{equation}
m = \frac{\pi - \alpha}{4 \pi G} \, ,
\end{equation}
and the total mass is 
\begin{equation}
M = \frac{n(\pi - \alpha)}{4 \pi G} \, .
\end{equation}
Taking a limit where the number of particles goes to infinity keeping the total mass $M$ fixed, each polygon becomes a disk of radius $R$ in hyperbolic space, with 
\begin{equation}
\label{ContinuumLimitShell}
\cosh R = 2 G M \; .
\end{equation}
The circumference of the disk is the limiting value of $C=nA$, which gives
\begin{equation}
C = 2 \pi \sinh R = 2 \pi \sqrt{(2 GM)^2 -1} \; .
\end{equation}

\subsubsection*{Boundary geometry}

\begin{figure}
    \centering
    \includegraphics[width = 0.5 \linewidth]{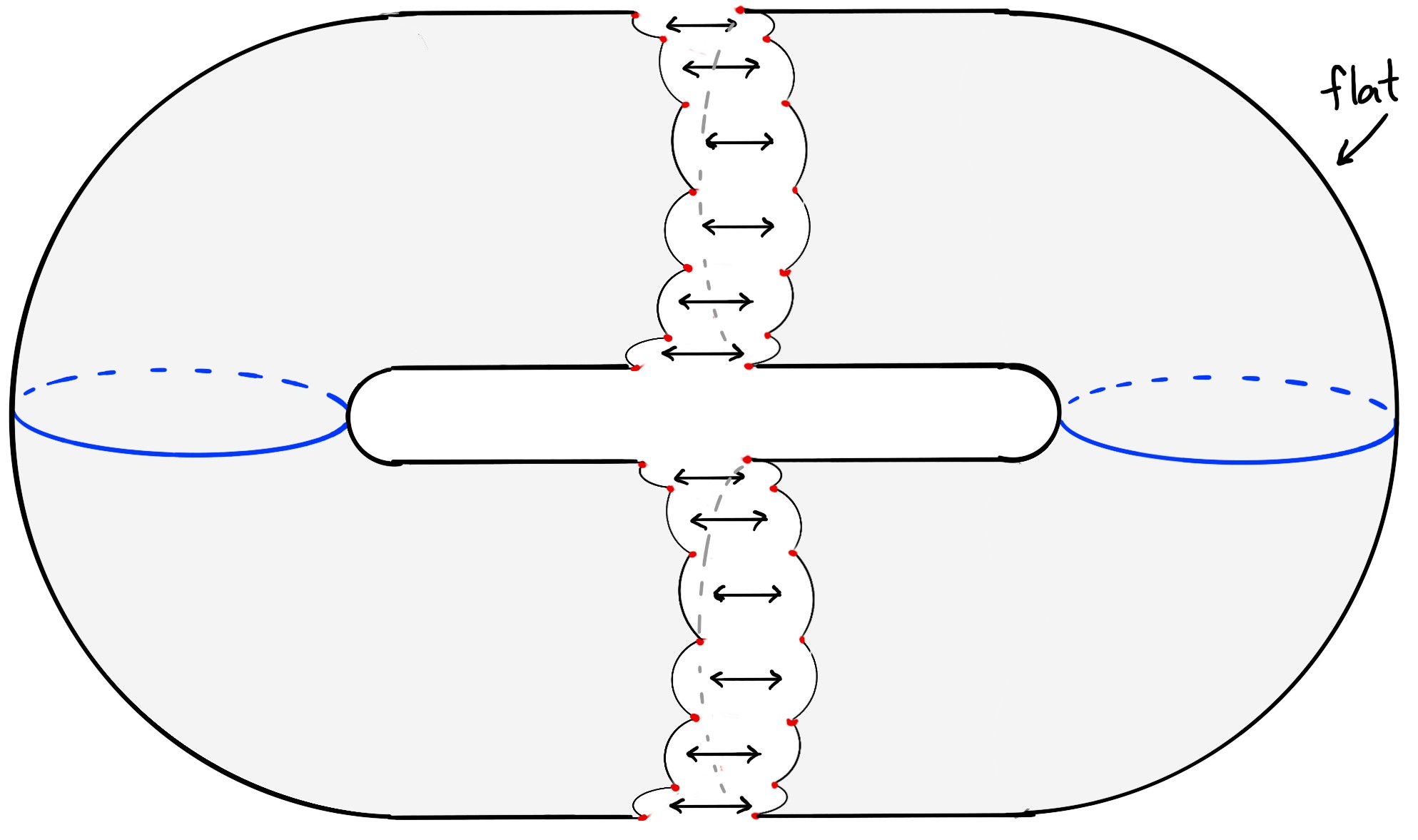}
    \caption{Boundary geometry for the AS${}^2$ construction of Figure \ref{fig:OG}. The geometry is flat away from the indicated identifications, which impart local curvature.}
    \label{fig:Boundary}
\end{figure}

The boundary of this AS${}^2$ construction is topologically a torus, but the boundary geometry as we have presented it so far is not flat because of the identifications, depicted in Figure \ref{fig:Boundary}. However, after a Weyl transformation, we can map it to a flat rectangular torus with two $\mathbb{Z}_n$-symmetric rings of operator insertions at opposite sides of the torus, as in the original AS${}^2$ construction.

\begin{figure}
    \centering
    \includegraphics[width = 0.4 \linewidth]{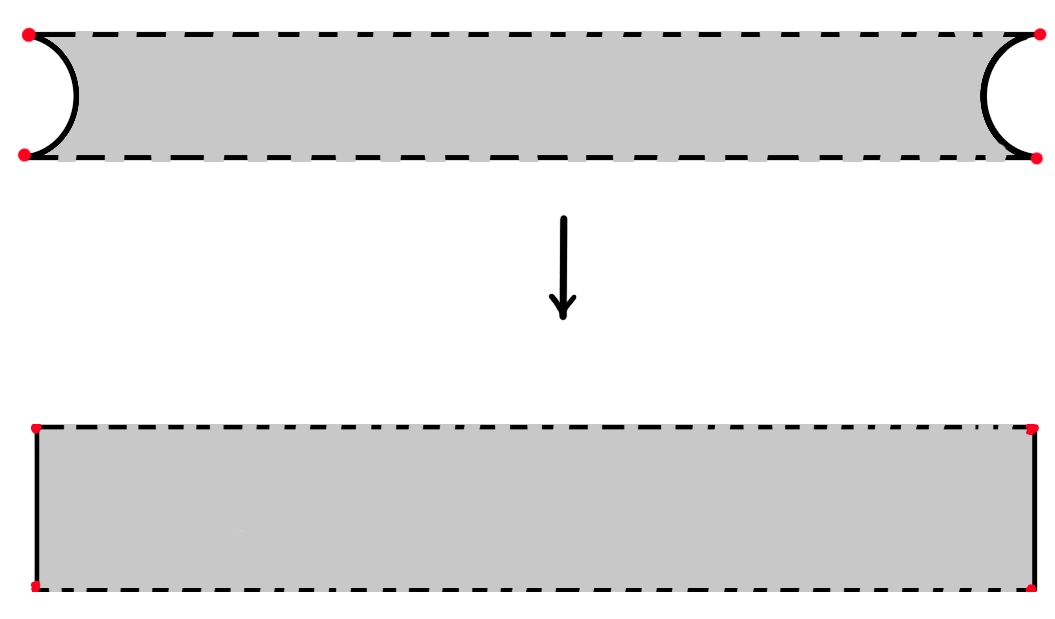}
    \caption{The boundary geometry in Figure \ref{fig:Boundary} may be divided into $2n$ equivalent flat pieces shown in the top figure. A Weyl transformation can map each of these pieces into a flat rectangle, after which the full geometry becomes a rectangular torus.}
    \label{fig:Mapping}
\end{figure}

To see this, we note that the boundary geometry divides into equivalent pieces as shown in Figure \ref{fig:Mapping}. Via a Weyl transformation, each of these can be mapped to a flat rectangle.\footnote{Here, we use the Riemann mapping theorem to guarantee a conformal transformation that takes the region to a disk with the corners mapping to four specific points around the boundary. There is  a further conformal transformation to take this to a rectangle with the corners of our original region mapping to the corners of the rectangle. The 2D conformal transformation is equivalent to a Weyl transformation plus a diffeomorphism.} Combining such Weyl transformations for all the elementary regions, we end up with a flat rectangular torus with the operator insertion in $\mathbb{Z}_n$ symmetric rings, as desired. 

The conformal field theories on these two geometries related by a Weyl transformation are equivalent, so our construction does give a saddle for the 2D CFT version of the original AS${}^2$ construction.

\subsection{Homogeneous and Isotropic AS${}^2$ geometries}

We now discuss a variety of other interesting examples that generalize AS${}^2$.

\subsubsection*{Approximately homogeneous and isotropic cosmologies}

With the general construction we have described, it is possible to find many examples where the cosmology is approximately homogeneous and isotropic, in contrast to the original AS${}^2$ where the matter is arranged on a single spherical shell. For example, we can start with any approximately homogeneous and isotropic triangulation of a unit sphere, replace all of the triangles with hyperbolic triangles whose lengths are any positive multiple of the side lengths in the triangulation, and then glue these together, adding particles of the appropriate masses. Each vertex in the sphere triangulation already has a deficit angle when the triangles are flat, and replacing with hyperbolic triangles with the same side lengths always decreases the angle, so we will always have positive deficit angles and positive masses in this construction. We can then make identifications on an arbitrary number of the AdS cylinders in the representation of Figure \ref{fig:GluedChewed} or add in an arbitrary number of tubes according to Figure \ref{fig:glueboth}.

\subsubsection*{Single AdS AS${}^2$ geometries}

As an interesting special case, we can have a large class of solutions where the associated AS${}^2$ geometry has only a single copy of AdS. For example, we can start with any polygon in hyperbolic space where the side lengths each have even multiplicity, and glue sides together in equal-length pairs. Provided that all the resulting vertices have conical deficits, we get a valid solution that involves only a single chewed AdS. The corresponding AS${}^2$ solution obtained by gluing the top and bottom parts of the AdS cylinder has only a single copy of AdS in the Lorentzian picture along with the cosmology. 

\subsubsection*{Vacuum AS${}^2$ solutions }

\begin{figure}[t!]
    \centering
    \includegraphics[scale=0.25]{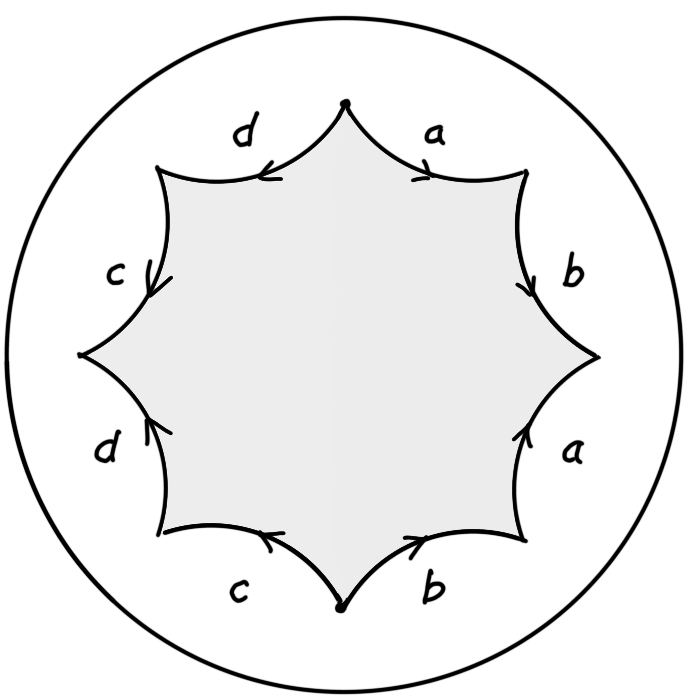}
    \caption{Identifying the indicated sides of a regular right-angled hyperbolic octagon gives a smooth genus 2 surface.}
    \label{fig:genus2}
\end{figure}

There are many cases where the surface $\Sigma$ is completely smooth, so we get a Maldacena-Maoz type cosmology (of spatial genus $g \ge 2$) without any matter. For the genus 2 case, we can start with a hyperbolic octagon with the identifications shown in Figure \ref{fig:genus2}. By making the analogous identifications in the corresponding chewed AdS, and also gluing the truncated top and bottom portions of the AdS cylinder, we get a family of AS${}^2$ solutions whose Lorentzian continuation has a single AdS together with the Maldacena-Maoz cosmology. An example of such a solution (with not one, but two AdS region) was discussed recently in \cite{Belin:2025ako}.

\begin{figure}[t!]
    \centering
    \includegraphics[width = 0.6 \linewidth]{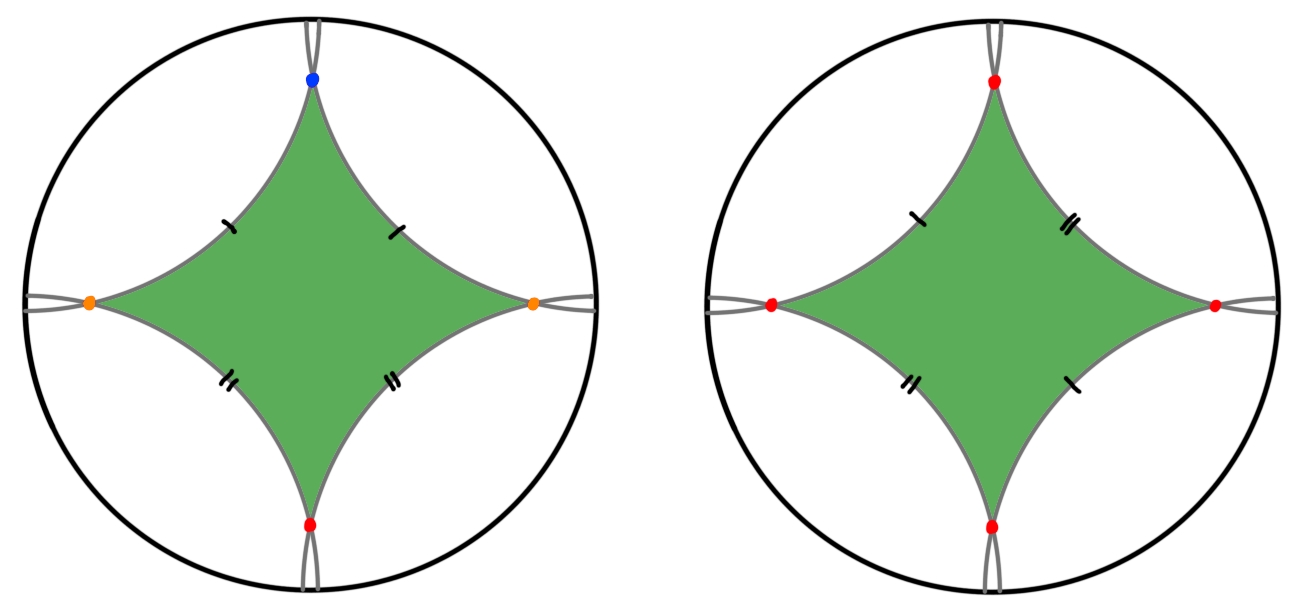}
    \caption{Self-gluings of a hyperbolic quadrilateral that give three particles on a sphere (left) or a single particle on a torus (right).}
    \label{fig:simple}
\end{figure}

\subsubsection*{One particle on a torus or three particles on a sphere}

The simplest examples with genus 1 and genus 0 have one and three particles respectively. We can obtain these from the identifications of a hyperbolic quadrilateral shown in Figure \ref{fig:simple}. Alternatively, we can make identifications starting with a pair of hyperbolic triangles.

\section{A necessary condition for dominance of the cosmological saddle}
\label{sec:condition}

In this section, we describe a test that can be used to evaluate whether a saddle of the gravitational path integral giving rise to a cosmology can be the dominant one.\footnote{This was mentioned briefly in footnote 2 of \cite{Sahu:2024ccg}.} 

\begin{figure}[h!]
    \centering
    \includegraphics[width=0.95\linewidth]{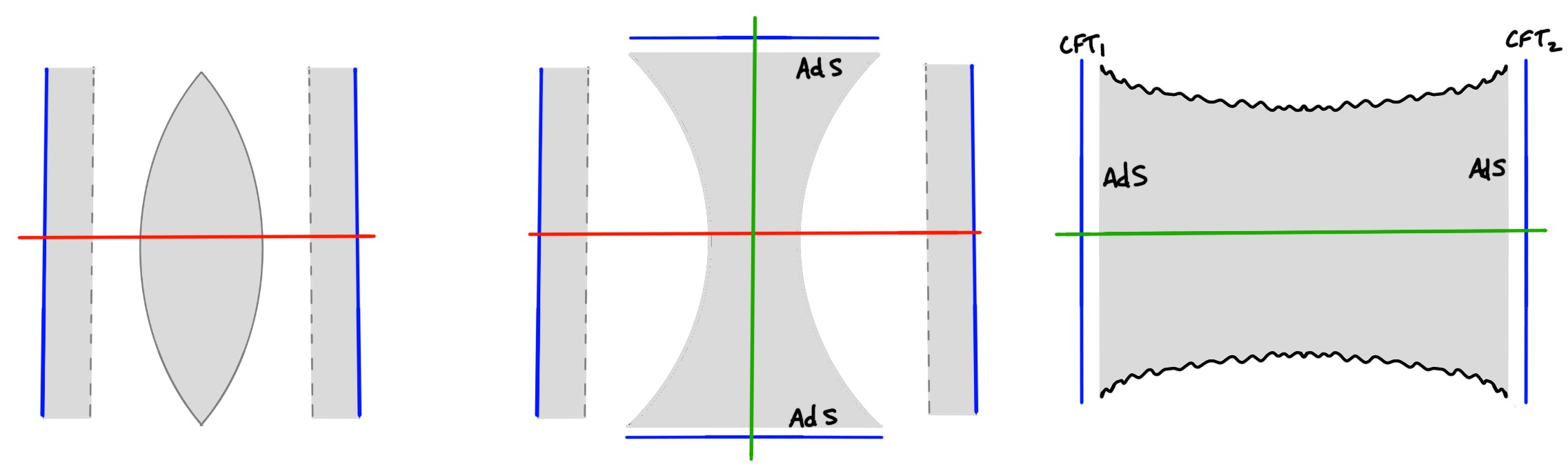}
    \caption{Middle: Saddle of a Euclidean gravity path integral that is associated with a path integral for some underlying holographic theory (blue).  The horizontal red slice gives initial data for a Lorentzian state whose evolution (left) includes a closed universe. The vertical green slice gives initial data for a two-sided back hole (right). If the Euclidean saddle giving the cosmology is dominant, the black hole on the right should be the dominant part of the wavefunction dual to the state of the two CFTs, which must then have order $c$ entanglement.}
    \label{fig:Slicings}
\end{figure}

The basic idea is the following: consider some Euclidean geometry that is proposed to be a dominant saddle for a path integral in a holographic theory. Suppose that some slicing of this gives the initial data for a closed universe cosmology. Now we instead consider an orthogonal slicing in a plane that includes the original Euclidean time. This slicing cuts through the length of the wormhole, so it gives initial data for a Lorentzian spacetime that includes a two-sided black hole -- this kind of slicing has been recently discussed in \cite{Belin:2025wju} for Maldacena-Maoz type wormholes. The existence of the black hole as the dominant part of the gravitational wavefunction implies that the dual CFT state has order $c$ entanglement (e.g., using the Ryu-Takayanagi formula). Thus, the Euclidean CFT path integral associated with the original Euclidean gravity path integral must be able to produce order $c$ entanglement between the two CFTs associated with the two sides of the wormhole. The various slicings described here are depicted in Figure \ref{fig:Slicings}.

\begin{figure}
    \centering
    \includegraphics[width = 0.5 \linewidth]{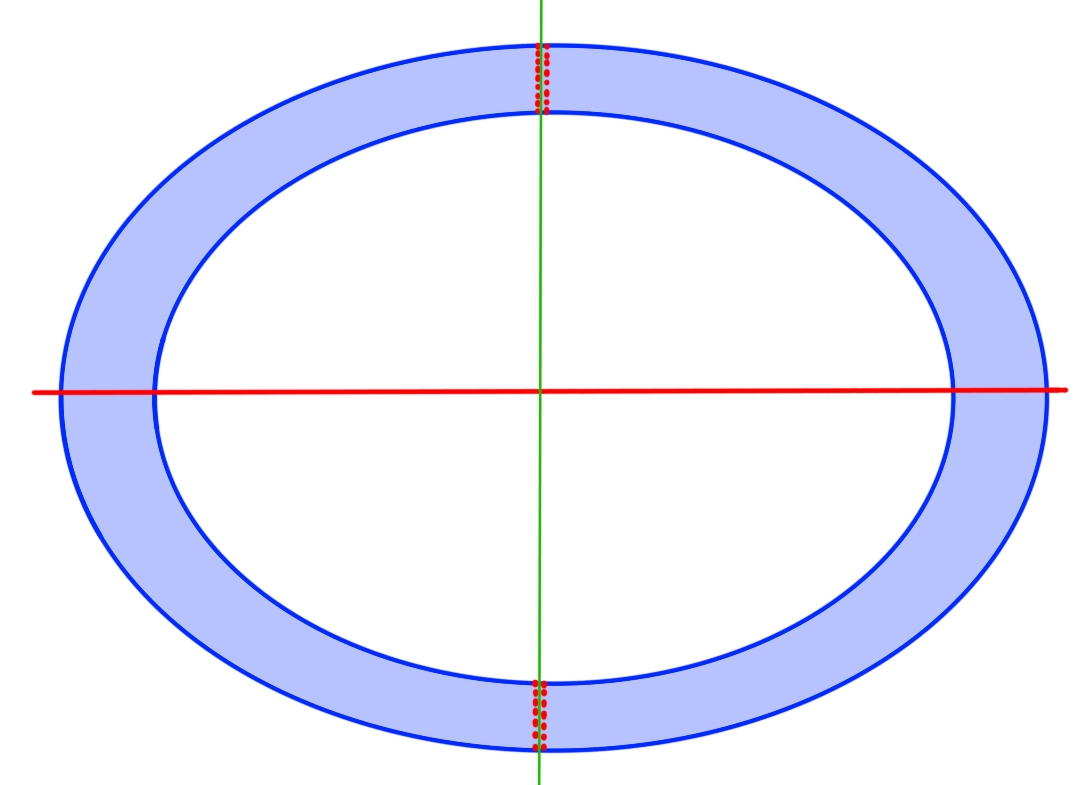}
    \caption{Two possible slicings for the Euclidean path integral in the AS${}^2$ construction (infinitesimally perturbed so that the operators are in two rings instead of one). If the horizontal slicing defines a state whose dual gravitational wavefunction is dominated by a spacetime with a baby universe, the vertical slicing should give a spacetime whose dual gravitational wavefunction is dominated by a two-sided black hole state. This requires entropy of order $c$ between the two CFTs, which is unlikely to exist for large $\beta$.}
    \label{fig:AS2slicings}
\end{figure}

\subsubsection*{Testing the original AS${}^2$ construction.}

\begin{figure}[t]
    \centering
    \includegraphics[width = 1.00 \linewidth]{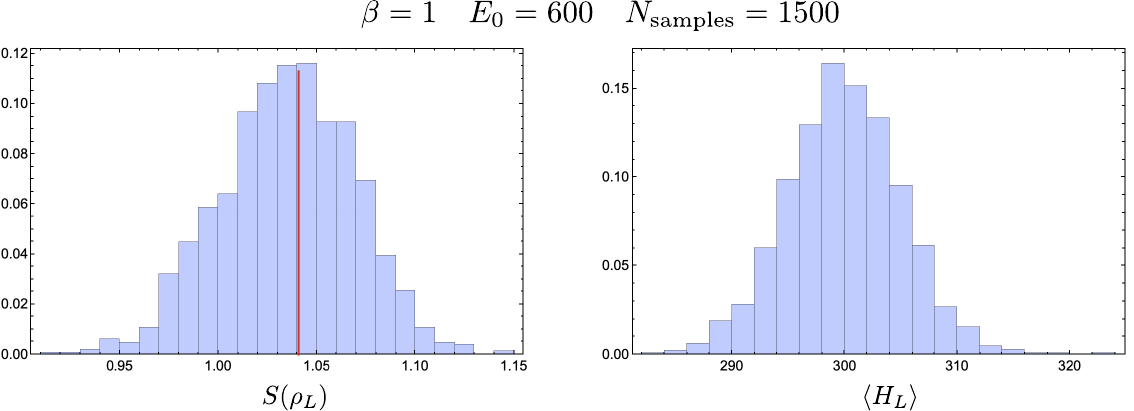}
    \caption{Relative frequency histograms showing the entanglement entropy (left) and average energy (right) of the left harmonic oscillator in the state \eqref{vertical}, for $N_{\rm samples} = 1500$ independent draws of the Hermitian operator $\mathcal{O}$ from a Gaussian distribution. We work in a $600$-dimensional truncation of the Hilbert spaces ($E_0 = 600$), and set units $\hbar \omega = 1$. The original TFD has $\beta = 1$, so that the initial entanglement entropy is $S_0 = 1.041$ (red vertical line on the left) and the initial left energy is $\langle H_L \rangle_0 = 1.082$. As explained in the text, $\mathcal{O}_L \otimes \mathcal{O}_R$ injects a large amount of energy without significantly changing the entanglement entropy.}
    \label{fig:HistogramsHarmonicOscillator}
\end{figure}

Let's apply this criterion to the original AS${}^2$ construction. The two relevant slicings are depicted in Figure \ref{fig:AS2slicings}. Here, we have infinitesimally adjusted the locations of the operator insertions so that half are on one side of the vertical green slicing and half are on the other side. 
Our criterion says that if the horizontal red slicing produces a state with a baby universe cosmology as the dominant contribution, the vertical green slicing should produce a state of two CFTs dual to a two-sided black hole and therefore having order $c$ entanglement. We can write this state explicitly as
\begin{equation}
\label{vertical}
|\Psi \rangle = {\cal O}_L \otimes {\cal O}_R | \Psi_{TFD}(\beta) \rangle \, ,
\end{equation}
where we start with the thermofield double state at large $\beta$ and then act with the same operator in each CFT. The operator ${\cal O}$ is defined by a path integral on an infinitesimal cylinder with a ring of operator insertions. This is not unitary, so the new state can have different entanglement from the original state, but in order to end up with order $c$ entanglement starting with order 1 entanglement would seem to require extreme fine tuning that essentially reverses the Euclidean evolution defining the thermofield double state. To check this, we performed a simulation with a pair of harmonic oscillators originally in the thermofield double state at some large $\beta$. Working in an $E < E_0$ truncation of each Hilbert space, we chose Gaussian random Hermitian operators $O$ (the norm of these does not affect the entanglement in the final normalized state, so we work at some fixed norm) and constructed the state \eqref{vertical} for each choice of $O$. This construction typically produced final states with energy of order $E_0$ but entropy of a similar order of magnitude to the original entropy, very small for large $\beta$. An example of such simulations can be found in Figure \ref{fig:HistogramsHarmonicOscillator}.

Our conclusion is that the original AS${}^2$ construction seems very unlikely to produce the cosmology as a dominant part of the wavefunction, since the construction did not involve any particular fine tuning such that the ring operators $O$ mimic backwards Euclidean evolution. In the CFT situation, we expect that the state \eqref{vertical} will have energy of order $c$ but entropy of order one, inconsistent with having a two-sided black hole as the dominant part of the dual gravitational state. From the bulk viewpoint, this argument indicates that there must be another saddle that dominates the computation of the norm of the state \eqref{vertical} and which does not include a cosmology. In the next section, we will discuss examples of such alternative saddles for the ${\rm AS}^2$ construction in three-dimensional gravity, but we emphasize that the previous argument is fully general, even if it does not tell us anything about what the alternative saddles are.

\subsubsection*{Can the baby universe ever dominate?}

Next, we ask whether the baby universe can ever dominate in any of our more general constructions. To satisfy our necessary condition, we would like a construction where in the dual slicing both the energy and the entropy are of order $c$.

Consider the situation where we have a sphere of total volume (i.e., 2D area) $V$ triangulated by a very large number $2n$ of hyperbolic triangles. We also have order $2n$ particles, so according to \eqref{Mass}, the mass of each particle is of order $V/(n G \ell^2)$. Now, consider the construction where we have a separate AdS cylinder for each triangle and we connect the top and bottom of each of these. Assume there is some spatial reflection symmetry so that there is a natural vertical slicing along the reflection-invariant slice such that $n$ of the cylinders coupling the two CFTs are on each side of this slice.

In this case, we have a state of the form
\begin{equation}
\label{verticalMultitube}
|\Psi \rangle = ({\cal O}_L \otimes {\cal O}_R) \otimes^n | \Psi_{TFD}(\beta) \rangle \, ,
\end{equation}
where now ${\cal O}$ is an operator that maps $n$ copies of the CFT Hilbert space to a single copy. Here the entropy is of order $n$, and again, we don't expect the operators to change the order of magnitude of the entropy. Thus, in order that the final state has order $c$ entanglement between the two CFTs, we would need $n$ to be of order $c$ and all the particle masses to be of order 1. This takes us outside the regime of validity of our classical construction (see \cite{Maloney:2025tnn} for a discussion of the quantum description that replaces it), but for large but finite $c$, we can certainly make the CFT construction. 

In summary, to have a chance to get the cosmology as a dominant saddle with tubes of length $\beta$ long enough so that the auxiliary AdS spacetimes do not contain black holes, we need the number of such tubes to be of order $c$. This means that most of the mass is in light particles and essentially every particle has order one entanglement with an auxiliary system. Note that in this case the boundary defines a CFT observable on a genus $c$ surface, and it is unclear to what extent a semiclassical analysis can be applied to the computation of such objects.\footnote{We thank Alex Belin for emphasizing this to us.} As a consequence, it is possible that this regime is also not able to make the cosmology the dominant saddle.

Alternatively, we can take sufficiently short tubes so that the auxiliary AdS spacetimes contain black holes; this gives the construction of \cite{Sahu:2024ccg}.

\subsubsection*{Additional averaging}

The discussion above assumed a fully microscopic construction with no additional ensemble averaging. It is well known that in some contexts, wormhole solutions compute certain CFT correlators averaged over an ensemble of theories \cite{Saad:2019lba, Almheiri:2019qdq,Penington:2019kki,Marolf:2020xie,Belin:2020hea,Maloney:2020nni,Chandra_2022,Chen:2020tes,VanRaamsdonk:2020tlr,Marolf:2021kjc,Saad:2021rcu,Cotler:2020ugk, Cotler:2021cqa,Balasubramanian:2022gmo, Pollack:2020gfa}. For three-dimensional geometries with matter insertions, there is a similar picture where the action for wormhole solutions matches with calculations in which CFT correlators are averaged over an ensemble of OPE data in the CFT \cite{Chandra_2022}. Such additional averaging applied to an AS${}^2$ construction built upon a given wormhole could also pick out the cosmological saddle. Failing our necessary condition can be taken as a sign that such additional averaging is required. In this case, knowing the details of this averaging would likely be required to understand better how much information the AS${}^2$ state (which will be mixed due to the averaging) carries about the cosmology.

\subsubsection*{Alternative saddles}

In cases where our necessary condition fails, we expect that the cosmological saddle cannot be dominant, so there should be some other saddle with lower action. In this other saddle, the asymptotic regions associated with the two CFTs in the alternative slicing should not be classically connected, so the wormhole should be absent in the Euclidean solution. In the next section, we discuss alternative saddles of this type that can be dominant. These alternative saddles can be understood as contributions to the CFT correlator where operators in a given ring contract among themselves.

\section{Other saddles}
\label{sec:othersaddles}

We have argued that in the standard  AS${}^2$ construction  described above, the cosmological saddle is unlikely to be the dominant one. In this section, we describe an alternative saddle (in the $D=3$ context) with the same boundary geometry that we expect to dominate in the limit of long narrow cylinders (large $\beta$ in the terminology of AS${}^2$). In a limit of small particle masses, we provide an explicit action calculation that verifies this dominance.

For now, we will assume that the number of insertions $n$ in each ring is even. Consider some region on the boundary with the topology of a disk and containing a pair of adjacent insertions. We will now describe the local geometry of a saddle where these two insertions pair up with each other. Consider Poincar\'e AdS${}_3$ with coordinates
\begin{equation}
\rd s^2 = \frac{\ell^2}{z^2}(\rd z^2 + \rd x^2 + \rd y^2) \, ,
\end{equation}
and the pair of geodesic hemispheres described by 
\begin{equation}
x^2 + (y \pm y_0)^2 + z^2 = 1 + y_0^2 \; .
\end{equation}
These intersect the boundary on a pair of circles which intersect at $y=0,x = \pm 1$. In the bulk, the hemispheres intersect on the geodesic
\begin{equation}
x^2 + z^2 = 1 \; .
\end{equation}
Now, we can remove from AdS${}_3$ the intersection of the two regions bounded by each hemisphere and the AdS boundary. The resulting cavity in AdS has two boundary components that are part of the two geodesic hemispheres. We can glue these together smoothly by identifying corresponding points related by the $y \to -y$ reflection symmetry. The resulting spacetime has a conical deficit along the geodesic $x^2 + z^2 = 1$ with angle
\begin{equation}
\tan \frac{\alpha}{2} = \frac{1}{y_0} \; .
\end{equation}
This corresponds to a particle of mass 
\begin{equation}
m = \frac{1}{4 \pi G} \arctan \frac{1}{y_0} \; .
\end{equation}

\begin{figure}
    \centering
    \includegraphics[width=0.5\linewidth]{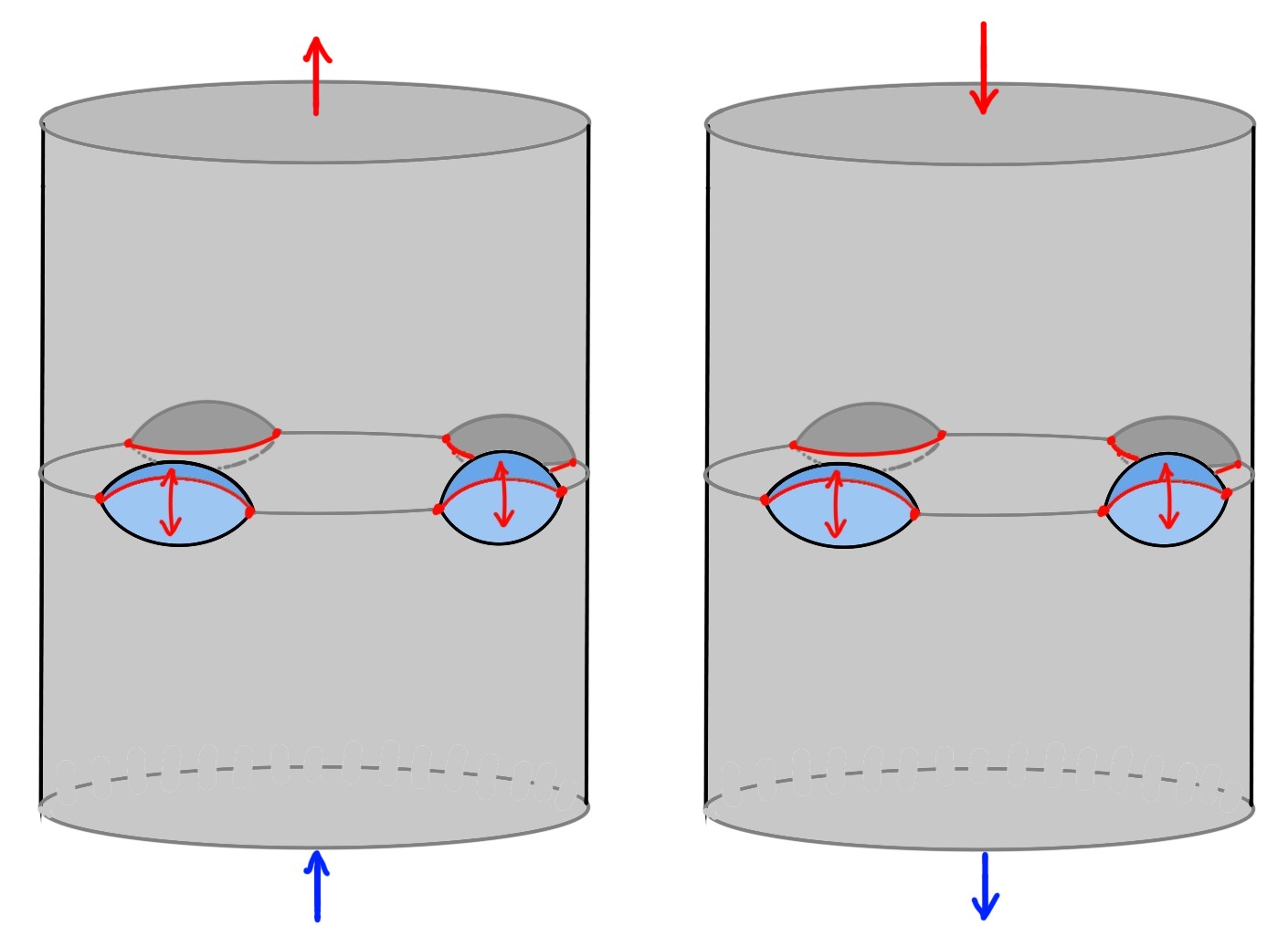}
    \caption{Alternative saddle in the 3D AS${}^2$ construction.}
    \label{fig:Removed}
\end{figure}

Now, we can map from Poincar\'e AdS back to the AdS cylinder in such a way that the boundary points $x = \pm 1, y=0$ map to points on the $\tau=0$ slice separated by some angle $\delta \theta$. As in the Poincar\'e description, the AdS${}_3$ has a portion removed with opposite sides identified to leave a conical defect connecting the two boundary points separated by $\delta \theta$. We can take $n/2$ copies of this, arranged symmetrically around the circle. Finally, we truncate the AdS cylinder in the past and future and glue it on to another copy. This is depicted in Figure \ref{fig:Removed}.

The resulting construction only has a $\mathbb{Z}_{n/2}$ symmetry. However, for some choice of the parameter $\delta \theta$, the conformal geometry of the boundary should be equivalent to the $\mathbb{Z}_n$ symmetric one in the AS${}^2$ construction above. 

To see this, note that there is a conformal transformation (Weyl transformation plus diffeomorphism) that takes the elementary region $R$ shown in Figure \ref{fig:Mapping} to a flat rectangle. Using this mapping and similar mappings for all the other copies of this elementary region, we can uniformize the boundary geometry to a flat torus in such a way that we preserve the $\mathbb{Z}_{n/2}$ symmetry, the reflection symmetry swapping the two copies that we have glued together, and the time-reflection symmetry. The final flat rectangular torus has two rings of $n$ points arranged in a $\mathbb{Z}_{n/2}$-symmetric way, with intervals between adjacent points that generally alternate between two different values $I_1$ and $I_2$ that depend on $\delta \theta$. For $\delta \theta$ going to 0 and $\infty$, the ratio of $I_1/I_2$ should take all possible values, so there exists some $\delta \theta_0$ where the ratio is 1 and we preserve $\mathbb{Z}_n$ symmetry.

We have thus described a second saddle for the ${\rm AS}^2$ CFT configuration. This depends on a choice of how to pair adjacent operators; by choosing the alternative pairing, we get a third saddle equivalent to this one.

\subsection{The dominant saddle}

We have presented two different bulk saddles for the ${\rm AS}^2$ CFT configuration: one including a closed cosmology in the bulk (which we will call the \emph{cosmological saddle}) and one in which contractions between matter particles are done locally in a single spatial cut and no closed cosmology exists in the time-reflection symmetric slice (the \emph{non-cosmological saddle}). We now want to compare their on-shell actions to see which one dominates in the limit of large $c$. In the general situation in which the particles produce a finite backreaction, obtaining the on-shell actions is non-trivial due to the conformal transformation needed in each case to make the boundary metric flat. However, we can look at a simplifying limit that takes us essentially to the original ${\rm AS}^2$ construction; in this limit, discussed already in \eqref{ContinuumLimitShell}, individual particles do not backreact and the cosmology is sustained by a large number of them, forming a continuous fluid of pressureless dust. More concretely, the limit is \cite{Anous:2016kss}
\begin{equation}
\label{DominantSaddles:AS2Limit}
1 \ll n \ll c = \frac{3 \ell}{2 G} \, , \qquad M = n m = \mathcal{O}\left(G^{-1}\right) \quad {\rm fixed} \, ,
\end{equation}
so that $1 \ll m \ell \ll c$. 

We first compute the Euclidean on-shell action of the cosmological saddle from
\begin{equation}
\label{DominantSaddles:EuclideanAction}
\mathcal{I} = - \frac{1}{16 \pi G} \int_{\mathcal{M}} \left( R + \frac{2}{\ell^2} \right) - \frac{1}{8 \pi G} \int_{\partial \mathcal{M}} \left( K - \frac{1}{\ell}\right) + \sum_{i=1}^n m_i \int_{\mathcal{W}_i} \rd L_i \, .
\end{equation}
The second term contributes only at asymptotic boundaries (where we set our boundary conditions to induce a flat metric), while the last term is the contribution from each point particle with worlidline $\mathcal{W}_i$ and length $L_i$. Away from the conical defects we have $R = - 6 / \ell^2$, so the first term gives a contribution proportional to the volume of the manifold. At the defects $R$ is singular; we can get the value of its integral by working in a local FRW patch
\begin{equation}
\rd s^2 = \rd u^2 + \cosh^2(u/ \ell) h_{ij} \rd y^i \rd y^j \, ,
\end{equation}
with $h_{ij} \rd y^i \rd y^j$ the metric of an infinitesimal hyperbolic disk $D_{\epsilon}$ with a conical defect $\delta$ at the origin. The Gauss-Bonnet theorem implies for the metric on a constant-$u$ slice
\begin{equation}
\int_{D_{\epsilon}} \rd ^2 y \sqrt{h} \, R[h] = 2 \delta \, ,
\end{equation}
as $\epsilon \to 0$, with $R[h]$ the scalar curvature of the metric $h$. For a three-dimensional infinitesimal tube around the defect $T_{\epsilon}$, and up to terms that vanish as $\epsilon \to 0$,
\begin{equation}
\int_{T_{\epsilon}} R = \int \rd u \int_{D_{\epsilon}} \rd ^2 y \sqrt{h} \, R[h] = 2 \delta \int_{\mathcal{W}} \rd u \, ,
\end{equation}
with the integral over the worldline $\mathcal{W}$ measuring the proper length of the particle trajectory. The singular curvature contribution from all the defects is then
\begin{equation}
- \frac{1}{16 \pi G} \sum_i \int_{T_{\epsilon, i}} R = - \sum_i  \frac{\delta_i}{8 \pi G} \int_{\mathcal{W}_i} \rd L_i \, .
\end{equation}
Using that $\delta_i = 8 \pi G m_i$, this exactly cancels the final term in \eqref{DominantSaddles:EuclideanAction}. We thus need to evaluate
\begin{equation}
\label{DominantSaddles:EuclideanActionSimplified}
\mathcal{I} = \frac{1}{4 \pi G \ell^2} {\rm Vol}[\mathcal{M}] - \frac{1}{8 \pi G} \int_{\partial \mathcal{M}} \left( K - \frac{1}{\ell}\right) \, .
\end{equation}

In order to induce the flat toroidal metric on the boundary, it is convenient to work with the metric of the AdS cylinder in the form
\begin{equation}
\rd s^2 = \ell^2\cosh^2(R/\ell) \rd \tau^2 + \rd R^2 + \ell^2\sinh^2(R/\ell) \rd \phi^2 \, ,
\end{equation}
with the asymptotic boundary at a large $R = R_{\infty}$. In the limit \eqref{DominantSaddles:AS2Limit}, the boundary insertions of Figure \ref{fig:OG} happen all at the same Euclidean time $\tau$, giving a spherically symmetric configuration. We can split the geometry to compute the on-shell action into three pieces: two long AdS tubes of size $\beta$ (as measured by $\tau$) and the region that arises from chewing and gluing the tubes and that includes the cosmological slice. Each of the AdS tubes contributes
\begin{equation}
\label{DominantSaddles:TubeAction}
\mathcal{I}_{\rm AdS}(\beta) = - \frac{\ell \beta}{8 G} \, ,
\end{equation}
with the boundary term in \eqref{DominantSaddles:EuclideanActionSimplified} regularizing the result. To obtain the contribution from the piece that results from gluing the chewed parts of the tubes, note that each particle follows a radial geodesic given by
\begin{equation}
\label{DominantSaddles:AS2Geodesics}
\tanh(R/\ell) = \frac{\sqrt{4 G^2 M^2 - 1}}{2 G M} \cosh(\tau) \, ,
\end{equation}
where we have set $\tau = 0$ at the turning point of the geodesics, shown in \eqref{ContinuumLimitShell} to be $\cosh(R/\ell) = 2 G M$. The geodesics have finite length in Euclidean time, $\tau \in (- \tau_{\infty}, \tau_{\infty})$ with
\begin{equation}
\tau_{\infty} = {\rm arccosh} \left( \frac{2 G M}{\sqrt{4 G^2 M^2 - 1}} \right) \, .
\end{equation}
Using the axial symmetry, the equation \eqref{DominantSaddles:AS2Geodesics} defines the surface of revolution traced by the cloud of particles. We can compute the action of this piece from the volume as
\begin{align}
\mathcal{I}_{\rm chewed} & = 2 \times \frac{1}{4 \pi G \ell^2} \int_0^{2 \pi} \rd \phi \int_{- \tau_{\infty}}^{\tau_{\infty}} \rd \tau \int_0^{R(\tau)} \rd R \, \ell^2 \cosh(R/\ell) \sinh(R/\ell) \\
\nonumber & = 2 M \ell \, {\rm arc cosh} \left( \frac{\sinh(R_{\infty}/\ell)}{\sqrt{4 G^2 M^2-1}} \right) - \frac{\ell}{G} \tau_{\infty} = 2 M \ell \log \left( \frac{e^{R_{\infty}/\ell}}{\sqrt{4 G^2 M^2 - 1}} \right) - \frac{\ell}{G} \tau_{\infty}  \, ,
\end{align}
where in the last step we have kept only non-vanishing terms as $R_{\infty} \to \infty$. The divergence we obtain is the standard one associated with $n$ boundary primary insertions of mass $m$; we could renormalize it away but we prefer to keep it to check that it also appears in the alternative saddle. Collecting the previous results, the on-shell action for the cosmological saddle is
\begin{equation}
\mathcal{I}_{\rm cosm.} = \frac{2c}{3} \left[- \frac{\beta}{4} - {\rm arccosh} \left( \frac{2 G M}{\sqrt{4 G^2 M^2 - 1}} \right) + 2 G M \log \left( \frac{e^{R_{\infty}/\ell}}{\sqrt{4 G^2 M^2 - 1}} \right) \right] \, ,
\end{equation}
with $c = 3 \ell / (2 G)$.

This is to be compared with the on-shell action of the non-cosmological saddle. In this case, the particles contract pairwise with each neighbor insertion, so that they do not create significant backreaction in the limit \eqref{DominantSaddles:AS2Limit}. We can treat them in a probe approximation, in which case the Euclidean action is that of an AdS tube of length $2 \beta$ plus a factor $m L$ for each particle, with $L$ the geodesic length between the insertions.\footnote{Here we are thinking of a full AdS tube with no conical defects at all, the contribution from the particles is included via the $m L$ factor. Alternatively, we could use the action \eqref{DominantSaddles:EuclideanActionSimplified} which makes no reference to the point particles, but then we would have to remove infinitesimal conical defects at the particle trajectories. Both perspectives are of course equivalent.} For the tube we can use the result \eqref{DominantSaddles:TubeAction} sending $\beta \to 2 \beta$. The length of the shortest geodesic between insertions separated by an angle $\Delta \phi = 2 \pi / n$ is (setting the regulating surface at $R = R_{\infty}$ again)
\begin{equation}
L = 2 \ell \log \left( \sin \left( \frac{\pi}{n} \right) e^{R_{\infty}/\ell} \right) \, .
\end{equation}
Note that $e^{-m L}$ with the divergence appropriately renormalized gives the conformal two-point function on a circle for $\Delta = m \ell \gg 1$, as it should. The total action is then, since we have $n$ particles and $M = m n$,
\begin{equation}
\mathcal{I}_{\rm non-cosm.} = \frac{2 c}{3} \left[ - \frac{\beta}{4} + 2 G M \log \left( \sin \left( \frac{\pi}{n} \right) e^{R_{\infty}/\ell} \right) \right] \, .
\end{equation}

We now compare the on-shell actions. Expanding at $n \gg 1$
\begin{equation}
\mathcal{I}_{\rm cosm.} - \mathcal{I}_{\rm non-cosm.} = \frac{2 c}{3} \left[ 2 G M \log \left( \frac{n}{\pi \sqrt{4 G^2 M^2 - 1}} \right) - {\rm arccosh} \left( \frac{2 G M}{\sqrt{4 G^2 M^2 - 1}} \right) \right] \, . 
\end{equation}
This difference is always large and positive in the regime we are working, given that our assumption $m \ell \ll c$ implies $n \gg G M$. We thus conclude that the non-cosmological saddle always dominates. In retrospect, this is the case because the length of a geodesic connecting the insertions is much shorter in the non-cosmological saddle than in the cosmological one:
\begin{equation}
L_{\rm cosm.} - L_{\rm non-cosm.} = 2 \ell \log \left( \frac{n}{\pi \sqrt{4 G^2 M^2 - 1}} \right) \, ,
\end{equation}
so that the cosmological saddle is suppressed by the matter propagators.

Some comments are in order regarding this computation. First, given two neighboring insertions in the non-cosmological saddle, there are infinitely many topologically inequivalent geodesics joining them (wrapping the thermal circle an arbitrary number of times). We have chosen the shortest (and thus dominant) one, but of course in principle one should sum over all possibilities, much like in a thermal two-point function. It is also important to note that, in the non-cosmological saddle, the choice of pairing for the boundary insertions is not unique. This is most obviously exemplified by the two alternative ways to pair neighboring operators, but in general there are also contractions in which non-neighboring insertions are paired.\footnote{The discussion in this section is in the limit in which the particles behave as non-backreacting probes. However, one could also try to build such saddles with a small number of particles with finite backreaction, via a slight generalization of the construction in Figure \ref{fig:Removed}.} In the probe limit with $n$ boundary insertions, we expect to have a number of contractions that grows exponentially with $n$.\footnote{For example, the number of ways to pair up $2k$ points on a circle without crossings is given by the Catalan number $C_k$, which has a large-$k$ asymptotics given by $C_k \sim 4^k / \sqrt{\pi k^3}$.} Regardless of the details of how these effects are accounted for, all of them will produce an enhancement of contributions to the wavefunction different from the one of the cosmological saddle, thus strengthening the conclusion that the cosmological saddle is subdominant.

\subsubsection*{Competing saddles in the thin-shell approximation}

As a check for the calculations in the previous section, we now describe another action comparison that makes use of the thin-shell approximation of \cite{Antonini:2023hdh}, but allows us to consider fully backreacted solutions for both saddles. We take a small perturbation of the AS${}^2$ construction, as described above, where the ring of operators is replaced with two parallel rings of operators separated by a small distance $\epsilon$. In this case, we have a version of the original AS${}^2$ saddle where the single matter shell supporting the wormhole is replaced by a pair of nearby shells. Each of these shells connects a ring of operator insertions with the corresponding ring related by the time-reflection symmetry. But we also have a ``local contraction'' saddle where a matter shell connects each ring of operator insertions with the nearby ring separated by $\epsilon$. These two saddles are depicted in Figure \ref{fig:ShellTwoSaddles}. The detailed comparison of the actions for these two saddles is described in Appendix \ref{App:SaddlesThinShell}. Again, we find that the cosmological saddle has larger action, with the action difference growing as $4c/3 \, G M \log (1/\epsilon)$ as $\epsilon \to 0$. In the previous calculation, the action difference grew as $4 c / 3 \, G M \log n$ for large $n$. This is essentially the same behavior, since $1/n$ and $\epsilon$ both play the role of determining the distance between the local contractions in the alternative saddle: between nearby operators in the first case and between nearby shells in the second case.

\begin{figure}[t]
    \centering
    \includegraphics[width = 0.95 \linewidth]{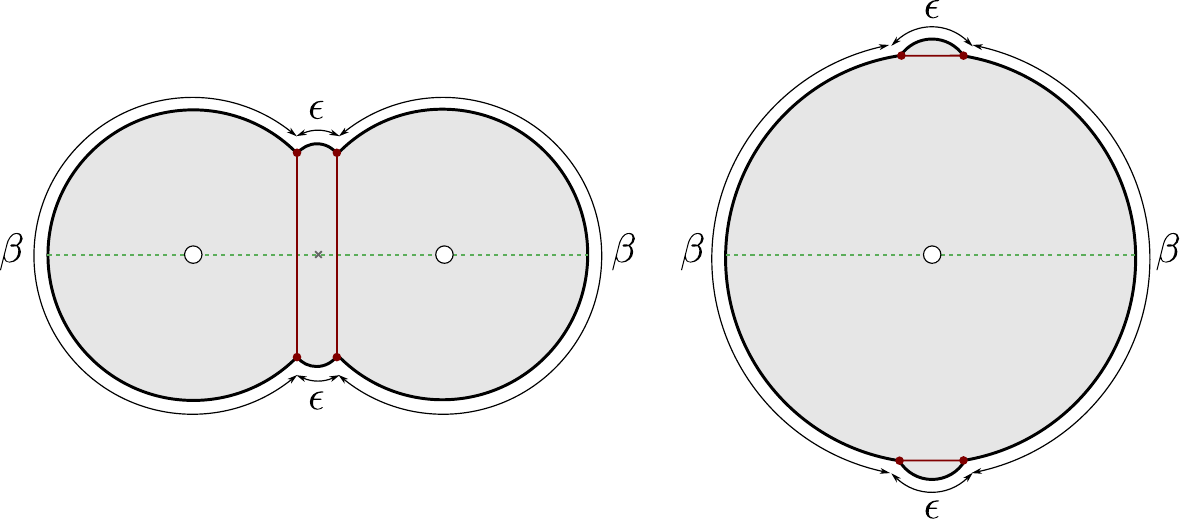}
    \caption{The two competing saddles when we split the ring of operators by two parallel rings separated a distance $\epsilon$. In each saddle, the dashed green line is the time-reflection symmetric slice on which we prepare the state. On the left, the slice includes a closed cosmology in the middle (white dots are not boundaries, but centers of spherical symmetry $r = 0$). The piece between the red thin shells is part of a Euclidean BTZ geometry. On the right, there is no closed cosmology, the state is prepared in two copies of AdS.}
    \label{fig:ShellTwoSaddles}
\end{figure}

\subsection{More general local saddles}

The existence of the saddle we have described relies on the neighboring operators being able to pair up; there should be a non-vanishing two-point function if only these operators are inserted into the path integral. The local geometry near a pair of operators in the full saddle is the same as the local geometry in the saddle that computes the two-point function. However, even if this two-point function vanishes, or we have an odd number of total operators, we expect that there would be similar saddles where collections of some small number $q > 2$ of neighboring operators with a non-vanishing $q$-point function ``contract'' with each other, such that the local geometry near these operator insertions in the full saddle matches with the local geometry in the saddle computing the non-vanishing $q$-point function. For all such saddles, the full geometry should only differ significantly from AdS${}_3$ in the asymptotic region near the rings of operator insertions. In the absence of symmetry reasons, having vanishing $q$-point functions for all local groupings of operators would likely require some extreme fine tuning in the choice of operators, if it is possible at all. This lines up with the general arguments in the previous section suggesting that the AS${}^2$ construction is not likely to have the cosmological saddle as the dominant one.

When the operator insertions creating the shell are charged under some symmetry we have to be more careful, since this can forbid the existence of a non-vanishing $q$-point function for all $q \leq n$. If the symmetry is an exact boundary global symmetry, dual to some bulk gauge symmetry, the cosmological saddle cannot exist: otherwise, in the Euclidean bulk we would have a closed spatial slice (the initial value surface of the cosmology) with non-vanishing gauge charge, violating Gauss's law.

We can however imagine that the particles going through the cosmology are charged under some flavor symmetry of the bulk EFT, which can be broken explicitly at some high-energy scale or by quantum gravity effects. In this situation, it is tempting to conclude that the dominant saddle would be the one in which charged operators in the ket are connected with the oppositely charged operators in the bra through the closed universe slice. However, note that the general argument presented in the previous section still applies, regardless of whether or not the particles are charged: for sufficiently large $\beta$, we do not expect the operator insertions to generate the $\mathcal{O}(c)$ entanglement needed to make the cosmological saddle the dominant one. One could imagine a competing non-cosmological configuration in which the  Euclidean spacetime is a solid torus and, possibly after some local interactions, the charge is transported from bra to ket via the lightest possible particles (to minimize the $e^{-mL}$ suppression). Clarifying whether this or some other type of saddle dominates over the cosmological one is a non-trivial an interesting problem but, as long as the general argument of the previous section applies, an alternative dominant saddle is expected to exist.

\section{Discussion}

In this paper, we have constructed a diverse menagerie of solutions generalizing the AS${}^2$ construction in three-dimensional gravity. These solutions each descend from a parent Euclidean wormhole solution via surgery where we connect that past and future conformal boundaries with tubes. These give rise to CFT boundaries on the time-reflection-symmetric slice, so there is an associated CFT state $|\Psi \rangle$ constructed by the CFT path integral.

A key question is to what extent this CFT state carries information about the cosmology. If the cosmological saddle provides the dominant saddle for the gravitational path integral with the specified boundary conditions, the gravitational wavefunction in the Lorentzian picture should be dominated by a spacetime that includes the cosmology. However, the necessary condition we have described (or the earlier general arguments in \cite{VanRaamsdonk:2020tlr}) suggest that this will only be the case if the Euclidean CFT construction leads to a large degree of correlation between the CFT degrees of freedom associated with the past and future boundaries of the cosmological wormhole. We have argued that the standard AS${}^2$ construction is not likely to provide the required amount of correlation. 

In order for the wormhole to dominate the path integral before adding the AS${}^2$, we require some type of ensemble or interaction that gives significant correlation between the CFTs (leading to order $c$ entanglement in the alternative slicing of section 3) \cite{VanRaamsdonk:2020tlr}. If this is not already present for the parent wormhole construction, adding a small number of thin CFT tubes connecting past and future will not be sufficient. If the added tubes are short and fat \cite{Sahu:2024ccg} or very numerous, they may on their own provide enough correlation to make the cosmological saddle dominate without additional ensemble averaging.

In cases where the cosmological saddle does not dominate, the cosmological spacetime is still expected to be present as a rare part of the wavefunction of the universe. It is an interesting question whether it is is possible to nevertheless extract the physics of the cosmology from this wavefunction, perhaps by projection on the CFT state \cite{Sahu:2024ccg} or through the recently described procedure of \cite{Belin:2025ako} (see also \cite{Belin:2025wju}) which replaces the pure state density operator with a particular mixed state one.

\subsubsection*{Cosmology from ``baryon'' asymmetry?}

It was suggested in \cite{Antonini:2023hdh} that a way to avoid non-cosmological saddles where the matter insertions self-contract within the individual rings would be to choose operators that insert particles with some global charge. In this case, at least naively, the insertions on one ring give rise to particle worldlines that must end on the other ring.

In quantum gravity, it is expected that there are no exact global symmetries (see e.g. \cite{harlow2021symmetries}), so the charge carried by any such particles can only be approximately conserved (as for baryon charge in the Standard Model). Thus, there should be some non-vanishing amplitude where we insert some number of these particles on on one side of the torus (the Euclidean past) and none on the other. Similarly, there could be an alternative saddle where the particles worldlines from the Euclidean past and future don't join up. The existence of these saddles require bulk interactions and so the saddles are suppressed at large $N$. But they can still be dominant if this suppression is less than the suppression (due to additional action) that comes from adding a wormhole part of the geometry for the charged particles to propagate through. It would be interesting to understand through a more detailed calculation whether inserting a large number of particles carrying some approximately conserved charge can cause the cosmological saddle to dominate. This would be fascinating, as it could imply that any cosmology produced by this construction exhibits something like baryon asymmetry. If this is the case, it is also possible that the explanation for baryon asymmetry in our actual universe could have a similar origin in quantum gravity.

\subsubsection*{Multi-universe saddles}

\begin{figure}
    \centering
    \includegraphics[width=0.9\linewidth]{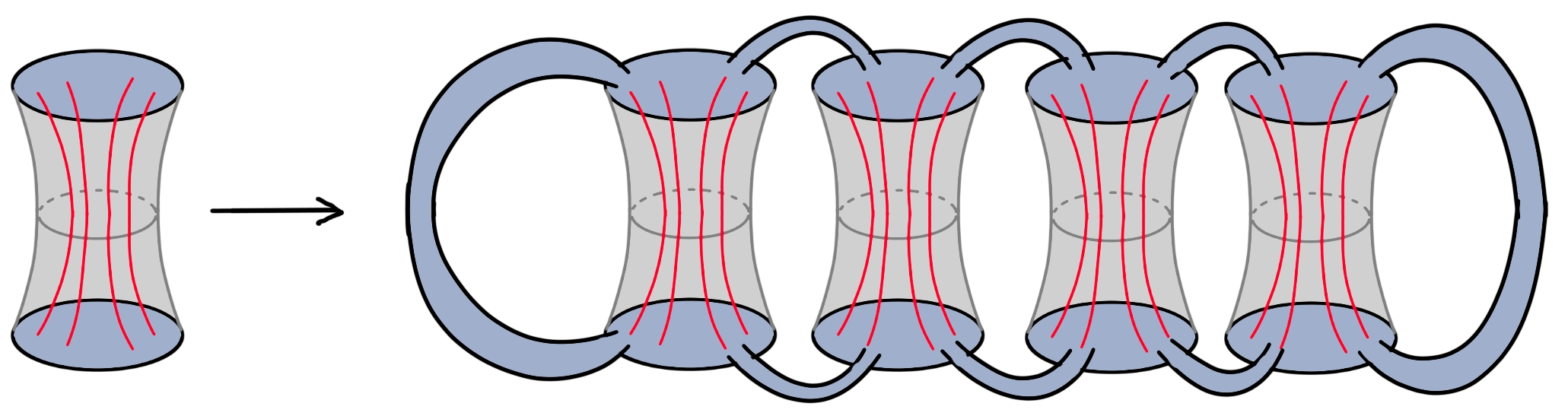}
    \caption{A multi-universe AS${}^2$ saddle constructed starting from four copies of a cosmological wormhole. Other saddles can be obtained by permuting the order that the wormhole tops connect to the wormhole bottoms.}
    \label{fig:Multi}
\end{figure}

The solutions that we have focused on in this paper have a single cosmological wormhole and a single baby-universe cosmology in the Lorentzian picture. But it is straightforward to generalize this to solutions with multiple disconnected cosmologies. Start with some number $n>1$ of basic cosmological wormholes as constructed in Figure \ref{fig:GluedChewed}. Each wormhole has a single connected boundary in the Euclidean past and a single connected boundary in the Euclidean future. Now we can perform gluing operations similar to the ones in Figure \ref{fig:gluing}. There, we connected the past and future boundaries of a single wormhole, but we can also connect the future boundaries of two different wormholes and the past boundaries of the same two wormholes in a way that preserves the time-reflection symmetry. More generally, we can connected any part of one boundary to any part of any other boundary, provided that we also add the gluing related by time-reflection symmetry. Through a combination of such gluings, we can produce many Euclidean solutions with a connected boundary whose Lorentzian continuations have multiple closed universes together with various copies of AdS. An example is shown in Figure \ref{fig:Multi}.

An interesting feature of the multi-universe saddles constructed from the same underlying wormhole is that we have a whole family of saddles with the same boundary geometry obtained by permuting how the wormhole bottoms connect to the wormhole tops. For saddles with a large number of universes, this can provide an exponential ($n!$) enhancement to the contribution of the cosmological saddle to the path integral. An interesting possibility is that this (or, more generally, having a large number of cosmological saddles with the same boundary through some other mechanism) might cause the set of cosmological saddles to dominate over non-cosmological saddles in the path integral overall even if non-cosmological saddles have lower action. In this case, the corresponding Lorentzian state would describe with high probability an entangled {\it polycosmos}\footnote{{\bf Polycosmos} (n.): a collection of mutually disconnected but quantum entangled cosmological spacetimes.} with a large number of disconnected closed universe cosmologies.

\subsubsection*{Implications for thin-shell black hole states}

The ${\rm AS}^2$ cosmologies supported by thin-shells were introduced as the low-temperature phase of a family of states labeled by $\beta$ \cite{Antonini:2023hdh}, collectively known as partially entangled thermal states \cite{Goel:2018ubv}. As we decrease $\beta$ (increase the temperature), the dominant saddle transitions to one in which the time-reflection symmetric slice is connected, forming a large Einstein-Rosen bridge that evolves in Lorentzian signature to a black hole with an expanded interior due to the presence of the matter shell. A family of such thin-shell black hole states were shown in \cite{Balasubramanian:2022gmo,Balasubramanian:2022lnw} to account for the microscopic entropy of Schwarzschild-like black holes, with later works extending them to a variety of other contexts \cite{Climent:2024trz,Balasubramanian:2024rek,Magan:2024aet,Balasubramanian:2025hns,Wang:2025jfd,Espindola:2025wjf,deBoer:2025tmh,Bavaro:2025ooe}.

Given that in this work we have argued that the cosmological saddle is generically not the dominant one in the ${\rm AS}^2$ configuration, is the same statement true when we lower $\beta$ and transition to the black hole phase? The first thing to notice is that the general argument in section \ref{sec:condition} ruling out the dominance of the cosmological saddle does not apply. Indeed, it is precisely as we lower $\beta$ enough to make the black hole phase to dominate that the dual two-sided state develops $\mathcal{O}(c)$ entanglement, compatible with a connected slice in both the horizontal and vertical slicings of Figure \ref{fig:Slicings}. Making more quantitative statements treating the particles forming the thin-shell individually would require a more careful analysis, but we can use a toy model analogous to the one in Appendix \ref{App:SaddlesThinShell} to estimate if the long wormhole saddle can dominate over the one in which the matter particles contract locally.

The long wormhole saddle is still penalized by the longer geodesics contracting between bra and ket, which give a factor $\sim c \, GM \log(n /(GM))$, where $1/n \sim \epsilon/\ell$ characterizes the separation between individual insertions. But now the on-shell action of a three-dimensional black hole is negative and inversely proportional to its inverse temperature, and the long wormhole phase contains two black holes of inverse temperature proportional to $\beta$, while the locally contracted saddle has a single black hole of inverse temperature proportional to $2 \beta$. These two contributions do not cancel (contrary to what happens in the cosmological phase), so the long wormhole is favored by a factor $\sim c /\beta$. As an order of magnitude estimate, we then expect that the long wormhole phase can dominate when
\begin{equation}
\frac{1}{\beta} \gtrsim GM \log \left( \frac{n}{GM} \right) \, .
\end{equation}
For large $n$, this requires a small $\beta$, but the dependence is only logarithmic.\footnote{Interestingly, if we make $GM$ very large, we seem to require an accordingly small $\beta$ to avoid the dominance of the locally connected saddle. It would be interesting to understand the implications of this fact for the heavy shell limit often taken in the black hole microstate counting works \cite{Balasubramanian:2022gmo,Balasubramanian:2022lnw,Climent:2024trz}.} It is then expected that, in the appropriate parameter range, the long wormhole saddle provides the dominant contribution, although a more detailed analysis than this order of magnitude estimate would be needed.

\section*{Acknowledgements}

We would like to thank Stefano Antonini, Alex Belin, Martin Sasieta and Brian Swingle for valuable discussions. This work is supported in part by the National Science and Engineering Research Council of Canada (NSERC) and the Simons Foundation via a Simons Investigator Award.

\appendix

\section{Competing saddles in the thin shell approximation}
\label{App:SaddlesThinShell}

In this appendix, we analyze in detail a toy model that shows why an alternative saddle to the ${\rm AS}^2$ cosmology can be expected to dominate. In the original ${\rm AS}^2$ construction, one considers the state
\begin{equation}
\ket{\Psi} = \sum_{ij} e^{- \beta_L E_i /2} \mathcal{S}_{ij} e^{- \beta_R E_j/2} \ket{i} \ket{j^*} \, ,
\end{equation}
on two copies of the CFT (with $\ket{j^{\star}}$ the CRT conjugate of $\ket{j}$). For simplicity, we only consider symmetric states $\beta_L = \beta_R \equiv \beta$, but the discussion can be generalized. The operator $\mathcal{S}$ creates a spherically symmetric thin shell of matter with total rest mass $M$, formed by a large number of point particles with $1 \ll m \ell \ll c$. The norm squared of the state is given by
\begin{equation}
\avg{\Psi | \Psi}= \tr\left(e^{- \beta H} \mathcal{S} e^{- \beta H} \mathcal{S} \right) \, ,
\end{equation}
where we consider uncharged shells with $\mathcal{S}^{\dagger} = \mathcal{S}$. In the ${\rm AS}^2$ construction, this norm is argued to be dominated by a gravitational saddle which contains a closed cosmology in its time-reflection symmetric slice. In this saddle, the two $\mathcal{S}$ insertions are connected through the closed cosmology in the bulk.

As argued in section \ref{sec:othersaddles}, we want to model an alternative saddle in which the particles within each $\mathcal{S}$ insertion annihilate among themselves. Instead of treating the particles individually, in this appendix we look at a simplified toy model, in which we imagine separating the particles in $\mathcal{S}$ in two equal groups, each with rest mass $M/2$ but inserted at slightly different Euclidean times (separated by $\epsilon$). In other words, we consider the state
\begin{equation}
\label{SaddlesThinShell:EpsilonState}
\ket{\Psi_{\epsilon}} = \sum_{ijk} e^{- \beta E_i /2} \tilde{\mathcal{S}}_{ik} e^{- \epsilon E_k} \tilde{\mathcal{S}}_{kj} e^{- \beta E_j/2} \ket{i} \ket{j^*} \, ,
\end{equation}
where $\tilde{\mathcal{S}}$ creates half of the shell inserted by $\mathcal{S}$. We will evaluate the norm squared $Z_{\epsilon} = \avg{\Psi_{\epsilon} | \Psi_{\epsilon}}$ using the bulk, in a saddle point approximation. One expects two contributions, depending on how the shells contract (see Figure \ref{fig:ShellTwoSaddles}). In one of them, both shells contract between bra and ket, and slicing through the horizontal time-reflection symmetric slice gives a state containing a closed cosmology. This cosmological saddle is essentially the ${\rm AS}^2$ one in the limit $\epsilon \to 0$. The other saddle contracts the shells within bra and ket, and slicing in the horizontal time-reflection symmetric surface simply gives a state of two entangled AdS regions. We will show that this non-cosmological saddle always dominates as $\epsilon \to 0$, and consequently the correct dual to \eqref{SaddlesThinShell:EpsilonState} for any non-zero $\epsilon$ does not contain a closed cosmology. The impossibility to find a non-cosmological saddle in the exact $\epsilon = 0$ limit is thus an effect of fine tuning together with the fact that the thin shell effective model of the matter does not allow the particles within each insertion to self-contract.

In the thin-shell approximation,\footnote{See \cite{Keranen:2015fqa,Anous:2016kss,Chandra:2022fwi,Sasieta:2022ksu} for relevant works using thin shells in contexts similar to the one of our discussion.} the bulk Euclidean action is 
\begin{equation}
\mathcal{I} = - \frac{1}{16 \pi G} \int_{\mathcal{M}} \left( R + \frac{2}{\ell^2} \right) - \frac{1}{8 \pi G} \int_{\partial \mathcal{M}} \left( K - \frac{1}{\ell}\right) + \int_{\mathcal{W}} \sigma \, ,
\end{equation}
where the final term is integrated over the worldvolume of the thin shells, $\mathcal{W}$, and $\sigma$ is their mass density. The field equations require the geometry to be locally hyperbolic away from the thin shells, while at the thin shells we get the junction conditions
\begin{equation}
h^+_{ab} = h^-_{ab} \equiv h_{ab} \, , \qquad k^+_{ab} + k^-_{ab} - h_{ab} \left(k^+ + k^- \right) = 8 \pi G \sigma u_a u_b \, .
\end{equation}
In these equations, we imagine the thin shell glues two pieces of the full manifold, $\mathcal{N}_{+}$ and $\mathcal{N}_{-}$. The first equation sets the induced metric to be the same from each side, while the second relates the energy in the shell to the discontinuity in extrinsic curvature $k^{\pm}_{ab}$. The normal to compute $k^{\pm}_{ab}$ is taken on both sides to be inward pointing.

Through any of the shells, we are gluing spherically symmetric solutions of the form
\begin{equation}
\rd s^2_{\pm} = f_{\pm}(r) \rd \tau_{\pm}^2 + \frac{\rd r^2}{f_{\pm}(r)} + r^2 \rd \phi^2 \, , \qquad f_{\pm}(r) = \frac{r^2}{\ell^2} - 8 G M_{\pm} \, ,
\end{equation}
where we chose the radial coordinate to match between both sides anticipating the continuous induced metric. AdS is obtained by taking $M_{\pm} = - 1 /(8 G)$, while for $M_{\pm} > 0$ we get BTZ solutions. The trajectory of the shell in the $(\tau_{\pm}, r)$ plane is parametrized as $\tau_{\pm} = \mathcal{T}_{\pm}(\lambda)$, $r = R(\lambda)$, and the induced metric is set to $h_{ab} \rd \xi^a \rd \xi^b = \rd \lambda^2 + R^2(\lambda) \rd \phi^2$ by choosing the parameter $\lambda$ satisfying
\begin{equation}
\label{SaddlesThinShell:EuclideanTimeDerivative}
\frac{\rd \mathcal{T}_{\pm}}{\rd \lambda} = \frac{1}{f_{\pm}(R)} \sqrt{f_{\pm}(R) - \dot{R}^2} \, ,
\end{equation}
with $\dot{R} = \rd R / \rd \lambda$. The junction conditions give
\begin{equation}
\kappa_- \sqrt{f_-(R) - \dot{R}^2} + \kappa_+ \sqrt{f_+(R) - \dot{R}^2} = 4 G \mu \, , \qquad \mu \equiv 2 \pi R \sigma(R) = {\rm const.} \, ,
\end{equation}
with $\mu$ the mass of the thin shell (we keep it general in this discussion, but for the constructions in Figure \ref{fig:ShellTwoSaddles} each shell has $\mu = M/2$). Note that $\kappa_{\pm}$ are signs (i.e., $\kappa_+ = \pm 1$, $\kappa_- = \pm 1$) that indicate the direction of the gluing. Namely, $\kappa = +1$ means we keep the part of the manifold inside the thin shell (i.e., with $r \leq R$) while $\kappa = - 1$ means we keep the part outside the thin shell ($r \geq R$). Squaring gives an effective radial equation independent of these signs,
\begin{equation}
\dot{R}^2 + V_{\rm eff} (R) = 0 \, , \qquad V_{\rm eff}(R) = - f_-(R) + \left( \frac{M_+ - M_-}{\mu} + 2 G \mu \right)^2 \, .
\end{equation}

The values of $\kappa_{\pm}$ are relevant once we insert $\dot{R}^2$ back into the junction condition
\begin{equation}
\kappa_- \left| \frac{M_+ - M_-}{\mu} + 2 G \mu \right| + \kappa_+ \left| \frac{M_+ - M_-}{\mu} - 2 G \mu \right| = 4 G \mu \, .
\end{equation}
Assume without loss of generality that $M_+ \geq M_-$ (and $\mu > 0$, since we consider positive mass shells). This equation forbids some types of gluings, such as $\kappa_+ =\kappa_- = -1$ and $\kappa_- = - 1 = - \kappa_+$; and sets bounds for others
\begin{equation}
\kappa_- = +1 = \kappa_+: \; 2 G \mu^2 \geq M_+ - M_- \, , \qquad \kappa_- = +1 = - \kappa_+: \; 2 G \mu^2 \leq M_+ - M_- \, .
\end{equation}

If we insert the shell at the asymptotic boundary, the effective potential determines the turning point $R_{\star}$ to be
\begin{equation}
\label{SaddlesThinShell:RStar}
\frac{R^2_{\star}}{\ell^2} = 8 G M_- + \left( \frac{M_+ - M_-}{\mu} + 2 G \mu \right)^2 \, ,
\end{equation}
in terms of which $V_{\rm eff}(R) = -(R^2 - R_{\star}^2)/\ell^2$. Using \eqref{SaddlesThinShell:EuclideanTimeDerivative}, we can also get the total Euclidean time elapsed by the trajectory
\begin{equation}
\Delta \tau_{\pm} = 2 \int_{R_{\star}}^{\infty}  \rd R \frac{\sqrt{f_{\pm}(R) + V_{\rm eff}(R)}}{f_{\pm}(R) \sqrt{-V_{\rm eff}(R)}} \, .
\end{equation}
The integral can be solved explicitly, the result is nicer if we treat separately the two types of backgrounds. For a black hole with mass $M_{\pm}$
\begin{equation}
\Delta \tau_{\pm} = \frac{2 \ell^2}{r_{\pm}} \arcsin \left( \frac{r_{\pm}}{R_{\star}} \right) = \frac{\beta_{\pm}}{\pi} \arcsin \left( \frac{r_{\pm}}{R_{\star}} \right) \, ,
\end{equation}
where $r_{\pm}$ and $\beta_{\pm}$ are the horizon radius (note that $\pm$ here refers to each side of the gluing, not inner and outer horizon) and inverse temperature of the black hole
\begin{equation}
\label{SaddlesThinShell:BlackHoleRelations}
r_{\pm}^2 = 8 G M_{\pm} \ell^2 \, , \qquad \beta_{\pm} = \frac{2 \pi \ell^2}{r_{\pm}} \, .
\end{equation}
For a shell propagating in the AdS background
\begin{equation}
\Delta \tau_{\pm} = 2 \ell \, {\rm arcsinh}\left( \frac{\ell}{R_{\star}} \right) \, .
\end{equation}

We now use these results to obtain the equations that determine the contribution from each saddle to $Z_{\epsilon} = \avg{\Psi_{\epsilon} | \Psi_{\epsilon}}$. Let us first look at the cosmological saddle, the left one in Figure \ref{fig:ShellTwoSaddles}. The parameters fixed by the boundary conditions are $\beta$, $\epsilon$ and the mass of each shell $M/2$; the saddle dynamically determines the inverse temperature of the AdS region $\beta_0$, the mass of the black hole $M_{\epsilon}$ (or its horizon radius $r_{\epsilon}$) and the turning point of the shell trajectory $R_{\star}$. The equations that implement this can be read from the previous results
\begin{subequations}
\begin{align}
\beta & = \beta_0 - 2 \ell \, {\rm arcsinh} (\ell/R_{\star}) \, , \\
\epsilon & = \frac{\pi \ell^2}{r_{\epsilon}} \left( 1 - \frac{2}{\pi} \arcsin(r_{\epsilon}/R_{\star}) \right) \, , \\
\nonumber \frac{R_{\star}^2}{\ell^2} & = - 1 + \frac{1}{(4 G M)^2} \left( \frac{r_{\epsilon}^2}{\ell^2} + 1 + 4 G^2 M^2 \right)^2 \\
& = \frac{r_{\epsilon}^2}{\ell^2} + \frac{1}{(4 G M)^2} \left( 4 G^2 M^2 -  \frac{r_{\epsilon}^2}{\ell^2} - 1 \right)^2 \, .
\end{align}
\end{subequations}
Note that, since we are gluing with $\kappa = + 1$ on both sides, we must respect the bound $4 G^2 M^2 \geq r_{\epsilon}^2 / \ell^2 + 1$.\footnote{This requires $2 G M \geq 1$. In the limit $\epsilon \to 0$, when we have a cosmology supported by a single shell of mass $M$, this bound is needed to have $R_{\star} \geq 0$ in \eqref{SaddlesThinShell:RStar}. Alternatively, from \eqref{Mass}, it is equivalent to the condition to have a cosmology supported by particles of total mass $M$ and with spherical topology ($\chi = 2$).}

We now obtain an analytic solution in the regime $\epsilon / \ell \to 0$. More precisely, we work in the regime in which $G M \epsilon/\ell \ll 1$ (given that $G M \geq 1/2$ cannot be arbitrarily small, this means $\epsilon / \ell \ll 1$ also). Since $r_{\epsilon}$ is upper-bounded in terms of $M$ in this gluing, the second equation above means that this regime must have $R_{\star}$ just slightly larger than $r_{\epsilon}$. This is what we would expect: the black hole piece in the saddle is very small, with the shell passing very close to its horizon. Explicitly, from the last two equations we find
\begin{subequations}
\label{SaddlesThinShell:SaddleCosmology}
\begin{align}
\frac{r_{\epsilon}}{\ell} & = \sqrt{4 G^2 M^2 - 1} \left( 1 - G M \frac{\epsilon}{\ell} + \frac{3}{2} G^2 M^2 \frac{\epsilon^2}{\ell^2} + \dots \right)  \, , \\
\frac{R_{\star}}{\ell} & = \sqrt{4 G^2 M^2 - 1} \left( 1 - G M \frac{\epsilon}{\ell} + 2 G^2 M^2 \frac{\epsilon^2}{\ell^2} \left( 1 - \frac{1}{16 G^2 M^2} \right) + \dots \right) \, .
\end{align}
\end{subequations}
Note how $R_{\star}$ indeed approaches $r_{\epsilon}$ as $G M \epsilon / \ell \to 0$. To have a large cosmology (large $R_{\star}/\ell$), we need to take also the limit of very heavy shells, $G M \gg 1$, but this was not assumed for the previous result.

\begin{figure}[t]
    \centering
    \includegraphics[width = 0.75 \linewidth]{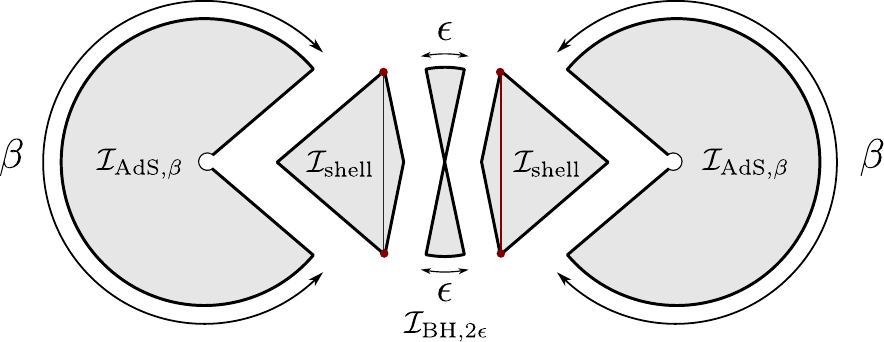}
    \caption{Decomposition of the saddle containing a closed cosmology.}
    \label{fig:SaddleCosmology}
\end{figure}

We may now evaluate the on-shell action of the solution. We decompose it as
\begin{equation}
\mathcal{I}_{\rm cosm.} = 2 \mathcal{I}_{\rm AdS, \beta} + \mathcal{I}_{\rm BH, 2 \epsilon} + 2 \mathcal{I}_{\rm shell} \, ,
\end{equation}
according to Figure \ref{fig:SaddleCosmology}. The first two terms are just pieces of thermal AdS or the BTZ black hole, their actions can be shown to be
\begin{equation}
\mathcal{I}_{\rm AdS, \beta} = - \frac{\beta}{8 G} \, , \qquad \mathcal{I}_{\rm BH, 2 \epsilon} = - \frac{\pi^2 \ell^2 \epsilon}{G \beta_{\epsilon}^2} = -\frac{\epsilon \left( 4 G^2 M^2 - 1 \right) }{4G} \, ,
\end{equation}
where in the second expression we keep just the leading order in $\epsilon$. The contribution from the diamond $X_{\rm shell}$ containing the shell can be obtained using the on-shell relation
\begin{equation}
R = - \frac{6}{\ell^2} + 16 \pi G \sigma \delta_{\mathcal{W}}(x) \, ,
\end{equation}
which shows that in this part of the geometry (where there is no Gibbons-Hawking-York or counterterm at $\partial \mathcal{M}$)
\begin{equation}
\mathcal{I}_{\rm shell} = - \frac{1}{16 \pi G} \int \left( R + \frac{2}{\ell^2} \right) + \int_{\mathcal{W}} \sigma = \frac{1}{4 \pi G \ell^2} {\rm Vol} \left( X_{\rm shell} \right) \, .
\end{equation}
The volume is evaluated by introducing a large radial cutoff $r_{\infty}$,
\begin{align}
\nonumber {\rm Vol}(X_{\rm shell}) & = 2 \pi \int_{R_{\star}}^{r_{\infty}} \rd R \frac{\sqrt{f_0(R) + V_{\rm eff}(R)}}{f_0(R) \sqrt{-V_{\rm eff}(R)}} R^2 + 2 \pi \int_{R_{\star}}^{r_{\infty}} \rd R \frac{\sqrt{f_{\epsilon}(R) + V_{\rm eff}(R)}}{f_{\epsilon}(R) \sqrt{-V_{\rm eff}(R)}} (R^2 - r_{\epsilon}^2) \\
 & = 4 \pi \ell^3 G M \log \left( 2 r_{\infty} / R_{\star} \right) - 2 \pi \ell^3 {\rm arcsinh} \left( \ell / R_{\star} \right) \, ,
\end{align}
where we have denoted by $f_0$ and $f_{\epsilon}$ the metric functions in the AdS and black hole regions, and we have kept only non-vanishing terms as we remove the regulator $r_{\infty} \to \infty$. We see then that the full result is
\begin{align}
\nonumber \mathcal{I}_{\rm cosm.} = & - \frac{c}{6} \left( \frac{\beta}{\ell} + \left( 4 G^2 M^2 - \frac{1}{2} \right) \frac{\epsilon}{\ell} + 4 \, {\rm arcsinh} \left( (4 G^2 M^2 -1)^{-1/2} \right) \right. \\
& \qquad \quad  \left. - 8 G M \log \left( \frac{2 r_{\infty} / \ell}{\sqrt{4 G^2 M^2 - 1}} \right) + \dots \right) \, ,
\end{align}
where we have used the form of the Brown-Henneaux central charge $c = 3 \ell / (2 G)$ and expanded $R_{\star}$ according to \eqref{SaddlesThinShell:SaddleCosmology}.

This is to be compared with the contribution from the non-cosmological saddle. Using hats to denote the parameters of this saddle, the equations are now
\begin{subequations}
\begin{align}
2 \beta & = \hat{\beta}_0 - 4 \ell \, {\rm arcsinh} (\ell/\hat{R}_{\star}) \, , \\
\epsilon & = \frac{2 \ell^2}{\hat{r}_{\epsilon}} \arcsin \left( \hat{r}_{\epsilon} / \hat{R}_{\star} \right)  \, , \\
\nonumber \frac{\hat{R}_{\star}^2}{\ell^2} & = - 1 + \frac{1}{(4 G M)^2} \left( \frac{\hat{r}_{\epsilon}^2}{\ell^2} + 1 + 4 G^2 M^2 \right)^2 \\
& = \frac{\hat{r}_{\epsilon}^2}{\ell^2} + \frac{1}{(4 G M)^2} \left( \frac{\hat{r}_{\epsilon}^2}{\ell^2} + 1 - 4 G^2 M^2 \right)^2 \, .
\end{align}
\end{subequations}
Note that we are now gluing in the exterior of the black hole region, so $\kappa_+ = - 1$ and $\kappa_- = + 1$. This implies that we have a lower bound on the mass of such black hole in terms of the mass of the shell, $\hat{r}_{\epsilon}^2 / \ell^2 + 1 \geq 4 G^2 M^2$.

As we send $\epsilon / \ell \to 0$ (compared to $G M$), the black hole in the saddle is expected to become very heavy compared with the shell. The last equation above then shows $\hat{R}_{\star} \gg \hat{r}_{\epsilon}$, so the turning point of the shell is actually very far from the horizon. We can solve in this regime, finding
\begin{subequations}
\label{SaddlesThinShell:SaddleNonCosmology}
\begin{align}
\frac{\hat{r}_{\epsilon}}{\ell} & = 4 \sqrt{\frac{\ell G M}{2\epsilon}} \left( 1 - \frac{G M}{12} \left( 1+ \frac{3}{4 G^2 M^2} \right) \frac{\epsilon}{\ell} + \dots \right)  \, , \\
\frac{\hat{R}_{\star}}{\ell} & = \frac{2\ell}{\epsilon} + \frac{2 G M}{3} + \frac{8 G^2 M^2 - 15}{180} \frac{\epsilon}{\ell} + \dots \, .
\end{align}
\end{subequations}

\begin{figure}[t]
    \centering
    \includegraphics[width = 0.75 \linewidth]{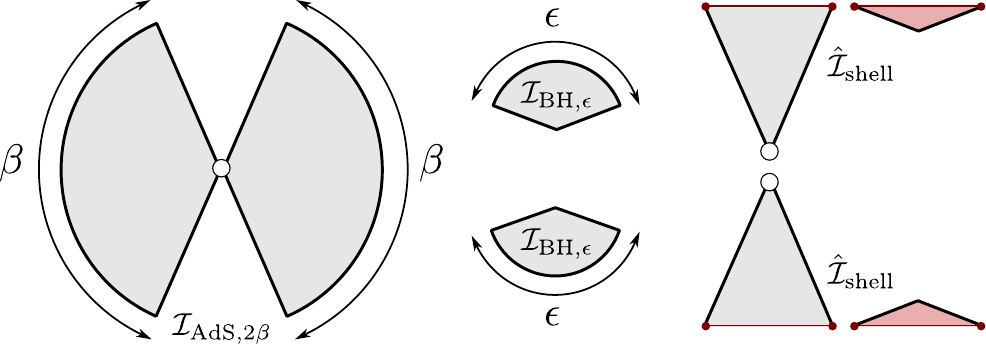}
    \caption{Decomposition of the non-cosmological saddle. On the rightmost piece, computing $\hat{\mathcal{I}}_{\rm shell}$, the red portion is to be subtracted from the gray one, since it was added to $\mathcal{I}_{\rm BH, \epsilon}$ without being part of the original saddle.}
    \label{fig:SaddleNoCosmology}
\end{figure}

It is convenient to evaluate the on-shell action by a method similar to the one used in the cosmological saddle. We view the saddle as a piece of thermal AdS, two full wedges of a black hole, and a triangle that ends in the shell and subtracts the unphysical part of the black hole wedge (see Figure \ref{fig:SaddleNoCosmology}). The on-shell action is decomposed as
\begin{equation}
\mathcal{I}_{\rm non-cosm.} = \mathcal{I}_{\rm AdS, 2 \beta} + 2 \mathcal{I}_{\rm BH, \epsilon} + 2 \hat{\mathcal{I}}_{\rm shell} \, ,
\end{equation}
The pieces without contributions from the shell are
\begin{equation}
\mathcal{I}_{\rm AdS, 2 \beta} = - \frac{\beta}{4 G} \, , \qquad \mathcal{I}_{\rm BH, \epsilon} = - \frac{\pi^2 \ell^2 \epsilon}{2 G \hat{\beta}_{\epsilon}^2} = - \ell M \, ,
\end{equation}
while the triangle containing the shell gives a contribution proportional to its volume, with
\begin{align}
\nonumber {\rm Vol}(\hat{X}_{\rm shell}) & = 2 \pi \int_{R_{\star}}^{r_{\infty}} \rd R \frac{\sqrt{f_0(R) + V_{\rm eff}(R)}}{f_0(R) \sqrt{-V_{\rm eff}(R)}} R^2 - 2 \pi \int_{R_{\star}}^{r_{\infty}} \rd R \frac{\sqrt{\hat{f}_{\epsilon}(R) + V_{\rm eff}(R)}}{\hat{f}_{\epsilon}(R) \sqrt{-V_{\rm eff}(R)}} (R^2 - \hat{r}_{\epsilon}^2) \\
 & = 4 \pi \ell^3 G M \log \left( 2 r_{\infty} / \hat{R}_{\star} \right) - 2 \pi \ell^3 {\rm arcsinh} \left( \ell / \hat{R}_{\star} \right) \, .
\end{align}
Combining all the terms, we get
\begin{equation}
\mathcal{I}_{\rm non-cosm.} = - \frac{c}{6} \left( \frac{\beta}{\ell} + 2 \frac{\epsilon}{\ell} - 8 G M \left( \log \left( r_{\infty} \epsilon / \ell^2 \right) - 1 \right) + \dots \right) \, ,
\end{equation}
where again we have expanded and kept only the leading pieces as $G M \epsilon / \ell \to 0$.

Comparing both results and neglecting terms that vanish as $\epsilon / \ell \to 0$, we get
\begin{equation}
\mathcal{I}_{\rm cosm.} - \mathcal{I}_{\rm non-cosm.} = \frac{2 c}{3} \left[ 2 G M \left( \log \left( \frac{2 \ell}{\epsilon \sqrt{4 G^2 M^2 - 1}} \right) + 1 \right) - {\rm arcsinh} \left( \frac{1}{\sqrt{4 G^2 M^2 - 1}} \right) \right] \, .
\end{equation}
As we take $G M \epsilon / \ell \to 0$, the logarithm becomes large, thus making the non-cosmological saddle to dominate.

As a final comment, in all the previous discussion we have assumed that we are in a regime in which the parameters defining the boundary conditions make the discussed saddles dominant over other existing ones. An example of such an extra saddle in the cosmological case would be one in which the middle part is another vacuum AdS geometry, instead of a black hole (cutting through the time-reflection symmetric slice, this saddle has two disconnected closed universes). For sufficiently large $\beta$ and sufficiently small $\epsilon$, this saddle is always subdominant.

\bibliographystyle{JHEP}
\bibliography{references}

\end{document}